\documentclass[journal=apchd5,manuscript=article]{achemso}

\usepackage{chemformula} % Formula subscripts using \ch{}
\usepackage[T1]{fontenc} % Use modern font encodings
\usepackage{comment}
\usepackage{multirow}
\usepackage{amssymb}
\usepackage{ulem}

\author{Emmanuel Lassalle}
\altaffiliation{These authors made equal contributions to this work.}
\author{Tobias W. W. Mass}
\altaffiliation{These authors made equal contributions to this work.}
\author{Damien Eschimese}
\altaffiliation{These authors made equal contributions to this work.}
\author{Anton V. Baranikov}
\altaffiliation{These authors made equal contributions to this work.}
\author{Egor Khaidarov}
\author{Shiqiang Li}
\author{Ramon Paniagua-Dominguez}
\email{ramon_paniagua@imre.a-star.edu.sg}
\author{Arseniy I. Kuznetsov}
\email{arseniy_kuznetsov@imre.a-star.edu.sg}
\affiliation{Institute of Materials Research and Engineering, A*STAR (Agency for Science, Technology and Research), 138634, Singapore}

\title[An \textsf{achemso} demo]
  {Imaging properties of large field-of-view quadratic metalenses and their applications to fingerprint detection}


\begin{document}

\begin{abstract}
Dielectric metasurfaces, extremely thin nanostructured dielectric surfaces, hold promise to replace conventional refractive optics, such as lenses, due to their high performance and compactness.
  However, designing large field-of-view (FOV) metalenses, which are of particular importance when imaging relatively big objects at short distances, remains one of the most critical challenges.
  Recently, metalenses implementing a quadratic phase profile
  have been put forward to solve this problem with a single element, but despite their theoretical ability to provide $180^\circ\,$FOV, imaging over very large FOV has not been demonstrated yet.
  In this work, we provide an in-depth analysis of the imaging properties of quadratic metalenses and, in particular, show that due to their intrinsic barrel distortion or fish-eye effect,
  there is a fundamental trade-off between the FOV achievable in a given imaging configuration and the optical resolution
  of the metalens and/or the detector resolution. To illustrate how to harness the full potential of quadratic metalenses, we apply these considerations to the fingerprint detection problem, and demonstrate experimentally
  the full imaging of a $5\,$mm fingerprint with features of the order of $100\,\mu$m, with a metalens ten times smaller in size and located at a distance of only $2.5\,$mm away from the object. This constitutes the most compact
  system reported so far for the fingerprint detection.
\end{abstract}

\section*{Introduction}

Modern electronic devices require increasing variety and number of functionalities while demanding ever more compact and lightweight designs.
Miniaturization of integrated systems fuel research and development of new concepts in \emph{e.g.} data storage, sensing and optics.
In recent years, flat optics emerged as a promising platform in which conventional refractive optical components are replaced by so called metasurfaces \cite{kildishev2013planar, yu2014flat, Kuznetsovaag2472}.
Metasurfaces are planar arrays consisting of numerous tailored nanostructures to orchestrate subwavelength light manipulation on a macroscopic scale. They can realize functionalities or devices
such as \emph{e.g.} polarization-sensitivity \cite{arbabi2015dielectric, yang2018generalized, Ramon2016},  waveplates \cite{kruk2016broadband, yu2012broadband}, holograms
\cite{ni2013metasurface, wang2016grayscale, huang2018metasurface, zheng2015metasurface, song2020largescale}, structural color printing
\cite{kumar2012printing, kristensen2016plasmonic, dong2017printing, flauraud2017silicon, proust2016all} and metalenses
\cite{arbabi2015subwavelength, khorasaninejad2016metalenses, lalanne2017metalenses, Khorasaninejadeaam8100}.
The latter are the flat optics analogue to conventional, bulk refractive lenses or lens systems, and can be tailored to address challenges related to \emph{e.g.} polarization-independent
imaging \cite{khorasaninejad2016polarization, arbabi2016multiwavelength}, high numerical aperture (NA) focusing \cite{paniagua2018metalens, arbabi2015subwavelength},
spherical aberrations correction \cite{aieta2013aberrations}, achromatic performance \cite{aieta2015multiwavelength, shrestha2018broadband, arbabi2017controlling, chen2019broadband, wang2018broadband}
and wide field-of-view (FOV) imaging \cite{pu2017, martins2020metalenses, engelberg2020near}. In particular, conventional wide field-of-view imaging requires lens systems consisting of multiple comparably
bulky and heavy refractive lenses \cite{samy2015simplified}, turning a flat optical solution particularly attractive.
In the past few years, flat optical approaches to wide FOV imaging included \emph{e.g.} aplanatic metasurfaces \cite{aieta2013aberrations}, metalens doublets \cite{arbabi2016miniature, groever2017meta} and numerically optimized phase profiles
\cite{kalvach2016aberration, shalaginov2020single, fan2020ultrawide}. Generally, the difficulty in realizing wide FOV imaging using a single metalens is to tailor simultaneously the optical
response of individual nanostructures as well as that of supercells to maintain the desired phase profile and avoid off-axis related abberations for a wide angular range. Recently, metalenses imparting a
quadratic phase profile have been proposed to achieve arbitrary wide FOV by trading off FOV against diffraction limited resolution equivalent to an NA of 0.27
\cite{pu2017, martins2020metalenses,   engelberg2020near}.

Here, we provide an in-depth analysis of the imaging properties of quadratic metalenses, identify their fundamental trade-offs and demonstrate unique application functionality for fingerprint detection. We confirm quadratic metalenses to be excellent flat optical solutions for wide FOV imaging but unveil that, in practice, the effective FOV is not arbitrary large as previously suggested \cite{martins2020metalenses}.
Our findings underline the importance of considering barrel distortion in combination with optical resolution limit and/or detector pixel size when designing quadratic metalenses for imaging configurations, especially for shorter distances.
After elaborating on the general imaging properties of quadratic lenses, we demonstrate short distance imaging of a fingerprint located in close proximity to the quadratic metalens,
a configuration suitable for portable electronic devices such as smartphones. Our results show that, using a single quadratic metalens, we are able to image a full fingerprint with a size of $5\,\text{mm}\times 5\,\text{mm}$ with a total
device thickness of only 2.7 mm. This is a considerable form-factor improvement compared to previous solutions, using \emph{e.g.} lens arrays to address this problem \cite{hu2020cmos}.

\begin{comment}
Here, we provide an in-depth analysis of the imaging properties of quadratic metalenses.
We first recall the focusing properties of quadratic metalenses and in particular their large-FOV ability, to subsequently explain how the imaging works.
We then present some considerations about the design of such metalenses, in particular related to the discretization of the phase, and fabricate a sample working in the near infrared (IR) that we experimentally and numerically
characterize.
Finally, we elaborate on the barrel distortion, or fish-eye effect, peculiar to large-FOV quadratic metalenses,
and highlight the existing trade-off between the FOV required to image an object and the image spatial resolution, given either by the detector or the lens, that in turn limits the capacity of imaging over arbitrary large-FOV. 
By apply these considerations to the particular problem of fingerprint capturing, we show that our fabricated quadratic metalens allows to image a full fingerprint using a single-lens system with a total thickness of $2.7\,$mm.
This is in contrast to previous propositions that used lens arrays in considerably thicker systems to address this problem \cite{hu2020cmos}. 
\end{comment}

\section*{Theory of quadratic metalenses}
\label{sec:theory}

\subsection*{Focusing properties}
\label{sec:focusing}

\begin{figure}[ht]
\centering\includegraphics[width=13cm]{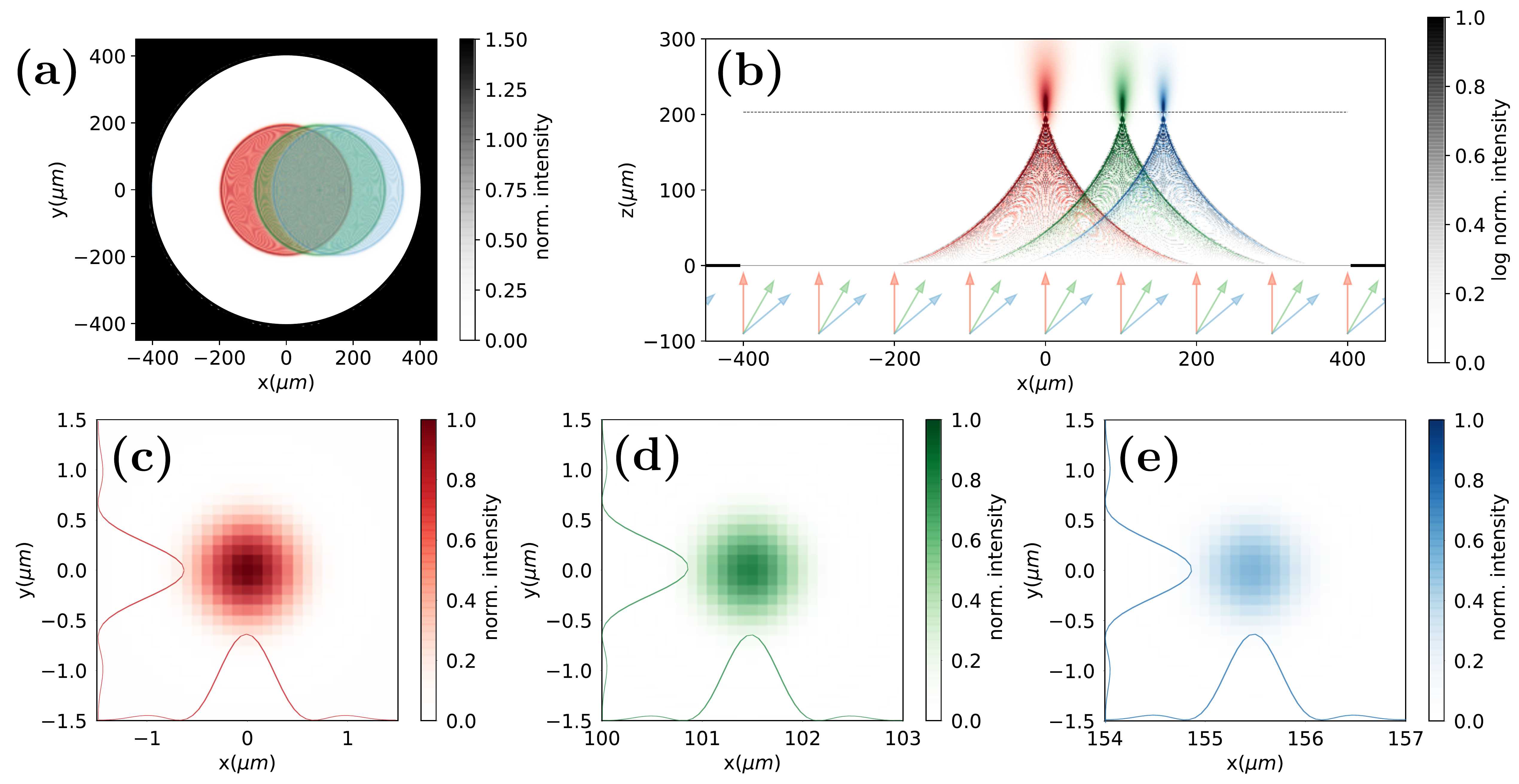}
\caption{Fourier optics simulations of the behaviour of a quadratic metalens of diameter $D=812\,\mu$m and focal length $f=203\,\mu$m, at the wavelength $\lambda_0=740\,$nm. (a) Cross-section in the metalens plane at $z=0$, for three different AOI: $\theta_\text{i}=0^{\circ{}}$, $30^{\circ{}}$ and $50^{\circ{}}$ in red, green and blue colours, respectively. The white area denotes the physical aperture of the metalens, and the coloured areas represent the effective areas for which the transmitted light is non-evanescent. (b) Propagation of an incident plane wave in $z$-direction through a quadratic metalens at $z=0$, same colour code. The focal plane is denoted by the black dashed line.}
  (c), (d) and (e) Cross-sections of the focal spots in the plane at $z=194\,\mu$m for the corresponding AOI, same colour code. The solid lines denote the intensity profiles along the $x$- and $y$- directions passing through the center of the focal spots.
  In all these color maps, the opacity levels are proportional to the normalized intensity and for (b) the normalized intensity is in log scale.
\label{fig:fig1}
\end{figure}

The general behavior of a phase-gradient metasurface is governed by the so-called ``generalized Snell's laws'', that can be derived from Fermat's principle \cite{genevet2017recent}.
For a metasurface working in transmission, this law states that an incident light impinging the metasurface at an angle $\theta_\text{i}$
is deflected by an angle $\theta_\text{t}$ given by \cite{yu2011light}:
\begin{equation}
n_\text{t}\,\text{sin}\left(\theta_\text{t}\right)=n_\text{i}\,\text{sin}\left(\theta_\text{i}\right)+\frac{\lambda_0}{2\pi}\frac{\partial \varphi}{\partial r}\; ,
\end{equation}
where $n_\text{i}$ ($n_\text{t}$) is the refractive index of the medium where the light is incident (transmitted), $\lambda_0$ is the wavelength of light in vacuum, $\varphi$
is the phase imparted by the metasurface, and $r$ denotes the position on the metasurface, which is centered at $r=0$.
This equation can be re-written as:
\begin{equation}
  k_\text{t}=k_\text{i}+\frac{\partial \varphi}{\partial r}\; ,
  \label{eq:wavevector}
\end{equation}
where $k_\text{i}\equiv k_0\,n_\text{i}\,\text{sin}(\theta_\text{i})$ [$k_\text{t}\equiv k_0\,n_\text{t}\,\text{sin}(\theta_\text{t})$]
is the projection of the incident (transmitted) wavevector on the metasurface plane and $k_0\equiv 2\pi/\lambda_0$ is the norm of the wavevector in vacuum.
In Eq.~(\ref{eq:wavevector}), the phase gradient $\partial \varphi/ \partial r$ can be physically interpreted ``as an effective wavevector, leading to a generalization of the conservation of the wavevector parallel to the surface''
\cite{genevet2017recent}.
A quadratic metalens is a metasurface that encodes the following quadratic phase profile \cite{pu2017, martins2020metalenses}:
\begin{equation}
  \varphi\left(r\right)=\varphi\left(0\right)-\frac{2\pi n_\text{t}}{\lambda_0}\frac{r^2}{2f}\; ,
  \label{eq:par_phase_profile}
\end{equation}
where $f$ is the focal length and the minus sign is for a converging lens.
Without loss of generality, we choose in this work the phase reference $\varphi\left(0\right)$ to be equal to $2\pi$,
but one can make any other choice of phase reference that may possibly depend on the wavelength.
%Interestingly, in Ref.~\cite{martins2020metalenses}, the authors observed that a metalens with such a quadratic phase profile corresponds to the limit of a spherical bulky lens with \emph{infinite} radius and refractive index.
Considering that only wavevectors $\left|k_\text{t}\right| \leq 2\pi n_\text{t}/\lambda_0$
propagate, this condition translates, in the case of the quadratic phase profile, into (by plugging Eq.~(\ref{eq:par_phase_profile}) into Eq.~(\ref{eq:wavevector})):
\begin{equation}
  \underbrace{f\left(\frac{n_\text{i}}{n_\text{t}}\,\text{sin}\left(\theta_\text{i}\right)-1\right)}_{r^\text{lower}(\theta_\text{i})}\leq r \leq \underbrace{f\left(\frac{n_\text{i}}{n_\text{t}}\,\text{sin}\left(\theta_\text{i}\right)+1\right)}_{r^\text{upper}(\theta_\text{i})}\; .
  \label{eq:evanescent2}
\end{equation}
This equation is illustrated in Fig.~\ref{fig:fig1}~(a) where we computed, based on Fourier optics calculations (see the Methods section), the propagation of a plane wave through a quadratic metalens for three different angles of incidence (AOI):
$\theta_\text{i}=0^{\circ{}}$, $30^{\circ{}}$ and $50^{\circ{}}$.
One can see in Fig.~\ref{fig:fig1}~(a) that the light is only transmitted through areas (in red, green and blue, respectively) that are smaller than the physical metalens aperture (in white). 
The parameters taken for the simulations are: wavelength $\lambda_0=740\,$nm, focal lens $f=203\,\mu$m and physical diameter $D=4f=812\,\mu$m, and considering air as the incident and transmitted medium ($n_\text{i}=n_\text{t}=1$). One can also see that the diameters of these effective areas are all equal to $2f$. Moreover, these areas are translationally shifted in the $x$-direction by the amount $f\,\text{sin}(\theta_\text{i})$, as the AOI $\theta_\text{i}$ increases from $0^\circ{}$ to $50^\circ{}$. Thus, as pointed out in \cite{pu2017}, the angular incidence is simply converted to a lateral shift of the phase profile.

\begin{comment}
As pointed out in \cite{pu2017}, one can see that the angular incidence is simply converted to a lateral shift of the phase profile by the amount $(n_\text{i}/n_\text{t})f\,\text{sin}(\theta_\text{i})$, thus maintaining the same focusing properties but displacing the focal spot by the same amount.
\end{comment}
We summarize hereafter the properties contained in the inequality of Eq.~(\ref{eq:evanescent2}), whom most of them have been presented in \cite{pu2017}. 
First, the area that transmits light, which we will refer to throughout this paper as the \emph{effective} working area of the metalens,
has a diameter defined as $D^\text{eff}\equiv r^\text{upper}(\theta_\text{i})-r^\text{lower}(\theta_\text{i})$ that reads:
\begin{equation}
  D^\text{eff}=2f\; .
  \label{eq:eff_diam_ideal}
\end{equation}
The previous equation implies that, even if the whole lens is illuminated, only a portion of it is actually transmitting the incident light.
This leads to an \emph{effective} numerical aperture defined as $\text{NA}^\text{eff}\equiv n_\text{t}\, R^\text{eff} /\sqrt{\left(R^\text{eff}\right)^2+f^2}$ that reads:
\begin{equation}
  \text{NA}^\text{eff}=n_\text{t} \frac{\sqrt{2}}{2}\; .
  \label{eq:eff_NA}
\end{equation}
Second, the effective working area is translationally shifted depending on the angle of incidence by the amount
$\Delta(\theta_\text{i})\equiv (r^\text{upper}(\theta_\text{i})+r^\text{lower}(\theta_\text{i}))/2$ which gives:
\begin{equation}
  \Delta(\theta_\text{i})=\frac{n_\text{i}}{n_\text{t}} f\text{sin}(\theta_\text{i})\; .
  \label{eq:delta_shift}
\end{equation}
Third, by considering the maximum positive angle of incidence possible, $\theta_\text{i} = 90^{\circ}$, we obtain the upper position
$r^\text{upper}$ which reads $r^\text{upper}=f\left(1+n_\text{i}/n_\text{t}\right)$. Conversely, by considering the maximum negative angle possible,
$\theta_\text{i} = -90^{\circ}$, we obtain the lower position $r^\text{lower}=-f\left(1+n_\text{i}/n_\text{t}\right)$.
This sets the maximum diameter of the lens that is active for the full range of angles of incidence, what we denote as the \emph{physical diameter} of the metalens $D\equiv r^\text{upper}-r^\text{lower}$, to:
\begin{equation}
  D=2f\left(1+\frac{n_\text{i}}{n_\text{t}}\right)\; ,
  \label{eq:phys_diam_ideal}
\end{equation}
which leads to the corresponding physical numerical aperture:
\begin{equation}
\text{NA}=\frac{n_\text{t}\left(n_\text{i}+n_\text{t}\right)}{\sqrt{n_\text{t}^2+\left(n_\text{i}+n_\text{t}\right)^2}}\; .
  \label{eq:phys_NA_ideal}
\end{equation}
We should stress here this important result. Namely, that there is no point in building a quadratic metalens with a diameter (and therefore NA) larger than the ones given by these equations.

\bigskip
In Fig.~\ref{fig:fig1}~(b), we show the propagation of the wave coming from $z<0$ through the quadratic metalens located in the plane $z=0$ keeping the same simulation parameters and considering the same three different AOI:
$\theta_\text{i}=0^{\circ{}}$, $30^{\circ{}}$ and $50^{\circ{}}$, as illustrated by the red, green and blue arrows, respectively.
One can see in all three cases that the light is focused around the focal plane at $z=f=203\mu$m (denoted by the black dashed line) without apparent distortion, demonstrating the remarkable focusing
performance of such metalenses for different AOI.
Interestingly, one can also observe an elongation of the focal spots across the $z$-direction, leading to a certain depth-of-focus (DOF), which is the region along the $z$-direction over which the focal spot remains tightly focused.
A further analysis of the wavefront produced by this metalens based on Zernike polynomials reveals that this elongation is mainly due to spherical and defocus aberrations
(see Supporting Information (SI) Section 1 and Figs.~S1 and S2).
This unveils the intrinsic non-stigmatic character of quadratic metalenses, in contrast with an ideal lens that focuses all incoming rays into a focal point.

The focal spot in the $xy$-plane is shown for the three cases in Figs.~\ref{fig:fig1}~red(c), (d) and (e), respectively.
Note that the plane considered here is where the maximum of intensity is found, at $z=194\,\mu$m, \emph{i.e.} slightly before the focal plane, which is an intrinsic property of this
type of lenses with spherical and defocus aberrations.
One can see that in all cases the focal spots have similar full width at half maximum (FWHM) of about $1\,\mu$m
in the $x$- and $y$-directions. This FWHM is about two times larger than that for a diffraction-limited lens with NA equal to $\text{NA}^\text{eff}\simeq 0.71$ [see Eq.~(\ref{eq:eff_NA})], which
stems precisely from the spherical aberrations mentioned above.
%From this, one can also deduce the DOF of such a metalens, by: $\text{DOF}\sim 4\,\text{FWHM}^2/\lambda_0=\lambda_0/\text{NA}^2\simeq 5.4\,\mu$m.

In order to be more quantitative, we calculated the focusing efficiency, defined as the power within the focal spot (integrated over the second Airy disk) divided by the power passing through the effective working area of diameter $D^\text{eff}=2f$, and the FWHM of the focal spot as a function of the AOI ranging from
$\theta^\text{min}_\text{i}=0^{\circ{}}$ to $\theta^\text{max}_\text{i}=85^{\circ{}}$.
The focusing efficiency and FWHM are found to be constant in the entire range of AOI, equal to $14\%$ and $1\,\mu$m, respectively (see SI Section~2 and Fig.~S3).
Therefore, the FOV, defined as two times the maximum AOI $\theta^\text{max}_\text{i}$ for which the metalens is able to focus an incoming light, is extremely large,
at least equal to $\text{FOV}\geq 2\,\theta^\text{max}_\text{i}=170^\circ{}$.
Also, the rather low efficiency can be explained by the fact that the energy is spread within the elongated focal spot, which is an intrinsic property of quadratic metalenses (note that if one considered the ratio of
the intensity in the focal spot divided by the intensity that impinges the total area of the metalens of physical diameter $D=4f$ instead, the focusing efficiency would be four times smaller).
We finally calculated the lateral displacement of the focal spot (see SI Section~2 and Fig.~S3) and found that it perfectly matches with the analytical formula of Eq.~(\ref{eq:delta_shift}),
which not surprisingly shows that the lateral displacement of the effective working area leads to a lateral displacement of the focal spot by the same amount.

\subsection*{Imaging properties}
\label{sec:imaging}

We now analyse the image formation with quadratic metalenses, based on ray-tracing methods. 
We apply the \emph{first-order} ray-tracing method, which consists in considering two particular rays coming from an object point to obtain the corresponding image point as the intersection of these two rays in the image space.
Note that, as we saw in Section~\ref{sec:focusing}, such metalenses are not stigmatic (\emph{i.e.}, they do not image a point source into a single image point) because of spherical aberrations, and therefore this approach will only be valid in the \emph{paraxial approximation} (that is for object points close to the OA and small AOI).

\begin{figure}[ht]
\centering\includegraphics[width=12cm]{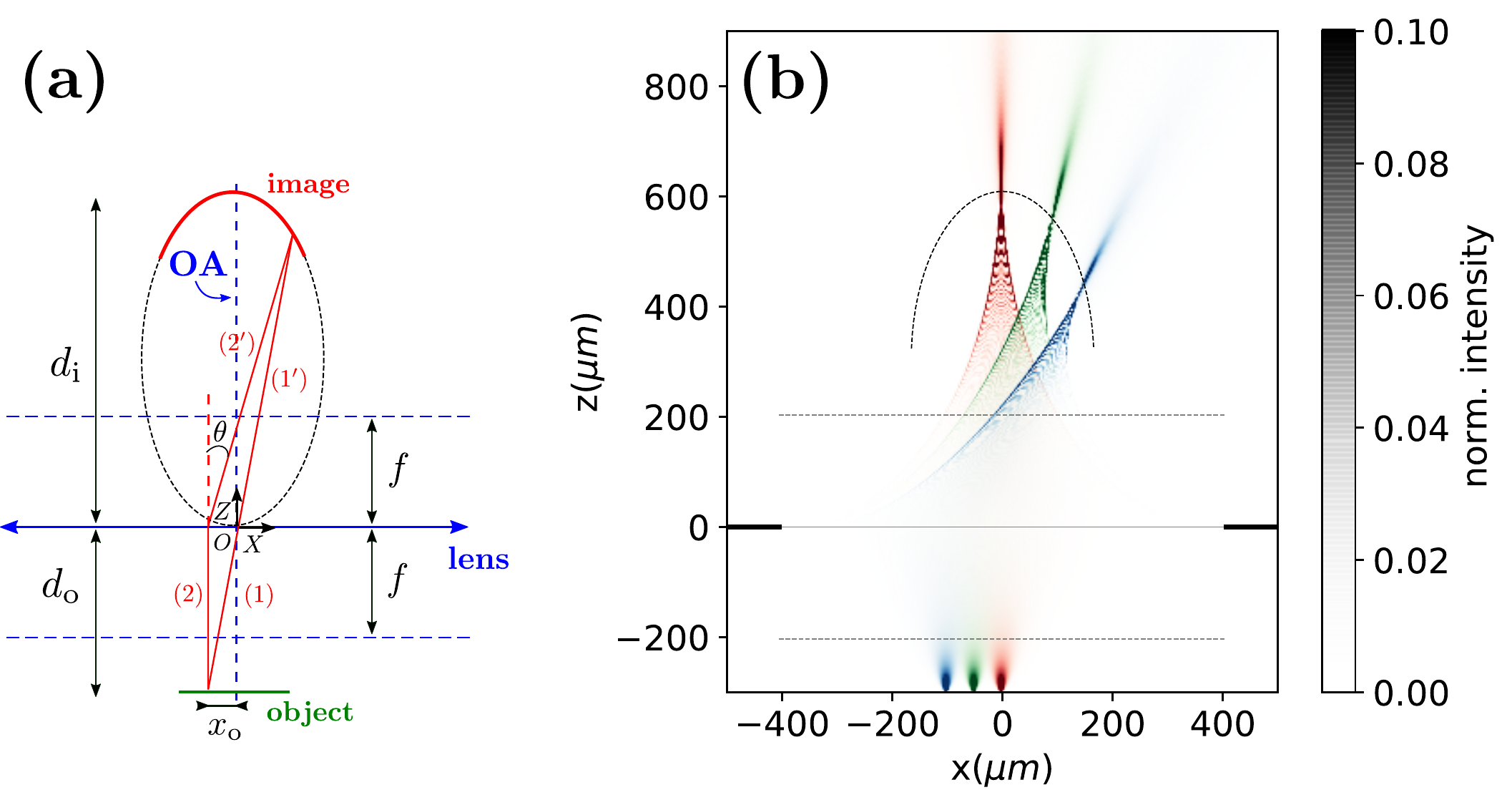}
\caption{(a) First-order ray-tracing for a quadratic lens. A quadratic lens (blue double-arrowed line) of focal length $f$ conjugates
  a planar object located at a distance $d_\text{o}$ (solid green line) into an image (solid red curve) which is formed along the ellipse given by Eq.~(\ref{eq:ellipse_image}) (black dashed curve), in the two dimensional Cartesian coordinate system with origin $O$ and axis lines $X$ and $Z$, oriented as shown by the arrows.
  In this configuration, $d_\text{o}=3f$ and $d_\text{i}=1.5f$. (b) Fourier optics simulations of the intensity maps obtained when imaging three points of a planar object in the same configuration as in (a).
Different colors correspond to different positions of the point in the object, with opacity levels being proportional to the normalized intensity. The ellipse given by Eq.~(\ref{eq:ellipse_image}) is also shown (black dashed curve).}
\label{fig:fig2}
\end{figure}

We thus consider two particular rays (straight red lines in Fig.~\ref{fig:fig2}~(a)), that come from the same point of an object located at a distance $x_\text{o}$ from the
OA: one ray passes the center of the metalens [labelled by $(1)$] and is not deflected when it emerges in the image space [labelled by $(1')$];
the other ray is parallel to the OA in the object space [labelled by $(2)$] and is deflected by the angle
$\theta$ according to $x_\text{o}=f\text{sin}\left(\theta\right)$ [Eq.~(\ref{eq:delta_shift})] in the image space [labelled by $(2')$].
It is this relation between the deflection angle and the distance to the center of the lens that makes this type of lenses behave completely different from ideal ones, for which an incident ray parallel to the OA passes the image focal point and that lead overall to a planar image with a certain magnification (see the SI Section~3 and Fig.~S4).
In particular, it implies that the intersection of the two rays $(1')$ and $(2')$ in the image space for different points in the object space obeys the following equation, in the Cartesian coordinate system of Fig.~\ref{fig:fig2}~(a) (see the SI Section~3 for derivation): 
%The intersection of the two rays $(1')$ and $(2')$ in the image space leads to the equation of an ellipse with width $2a_\text{i}$ and height $2b_\text{i}$
%(see details of the derivation in the \textcolor{red}{SI Section~2}).
%However, in the case of object points very far from the OA, the rays will asymptotically be parallel to each other
%and reach the lens with an AOI close to $\theta_\text{i}=90^\circ{}$, so that they will be focused in the image focal plane and at a distance $f$ from the OA,
%according to Eq.~(\ref{eq:delta_shift}). %and as seen in Fig.~\ref{fig:fig1}~\textcolor{red}{(a)} for different AOI.
%Thus, the previous previous ellipse equation must be amended by imposing that the point $(f,f)$ is part of the ellipse, which leads to the new $b_\text{i}$ factor:
\begin{equation}
\frac{(z-z_0)^2}{a^2_\text{i}}+\frac{x^2}{b^2_\text{i}}=1\quad\text{with}\quad \left\{
    \begin{array}{ll}
       z_0=\frac{1}{2}\frac{d_\text{o}f}{d_\text{o}-f}  \\[2mm]
       a_\text{i}=\frac{1}{2}\frac{d_\text{o}f}{d_\text{o}-f} \\[2mm]
       b_\text{i}=\frac{f^2}{\sqrt{2d_\text{o}(d_\text{o}-f)}}
    \end{array}
    \right.\; ,
    \label{eq:ellipse_image}
\end{equation}
where $f$ is the focal length and $d_\text{o}$ the distance between the object and the lens (all quantities being taken as algebraic quantities). Eq.~(\ref{eq:ellipse_image}) describes an ellipse centered on $(0,z_0)$ with width $2a_\text{i}$ and height $2b_\text{i}$.
Therefore, a quadratic metalens conjugates a planar object (green line in Fig.~\ref{fig:fig2}~(a)) into an elliptical image (red curve in Fig.~\ref{fig:fig2}~(a)). Note that this is valid for an object located at a distance $d_\text{o}>f$, and for the case of an object located between the focal plane and the lens $d_\text{o}<f$, the metalens creates a virtual image in the object space that is described by an hyperbola equation (see SI Section 3). This result, which is one of the key imaging properties of this type of lenses, can be generalized to the case of a 2D planar object: the ellipse (hyperbola) becomes an ellipsoid (hyperboloid) with same $b_\text{i}$ factor for the $x$ and $y$ coordinates. 

To verify this, we computed with Fourier optics simulations in Fig.~\ref{fig:fig2}~(b) the imaging by a quadratic lens of three different points of a planar object located at a distance $d_\text{o}=1.5f$ from the lens.
One can see that the corresponding image points follow the ellipse given by Eq.~(\ref{eq:ellipse_image}) (dashed black line in Fig.~\ref{fig:fig2}~(b)).  

Another interesting result is obtained when one considers the situation in which the distance to the object is much larger than the focal length of the lens ($d_\text{o}\gg f$).
In this case, usually met when imaging objects that are located very far away from the lens,
%On the other hand, one can see from Eq.~(\ref{eq:ellipse_image}) that if the condition $d_\text{o}\gg f$ is met,
a quadratic metalens can image a \emph{planar} object into a \emph{planar} image,
which is usually what one wants to achieve in practice.
Indeed, to first order in $f/d_\text{o}$,
the ellipse center becomes $z_0\rightarrow 0.5 f$ and its axis $a_\text{i}\rightarrow 0.5 f$ so that the image is formed in the image focal plane (located at a distance $f$ from the lens).
Also, for object points very far from the OA (for which the previous analytical approach fails), the rays will asymptotically be parallel to each other and reach the lens with an AOI close to $\theta_\text{i}=90^\circ{}$, and we know, following Eq.~(\ref{eq:delta_shift}), that these points will be focused in the image focal plane and at a distance $f$ from the OA.
Therefore, in the case of far-way objects ($d_\text{o}\gg f$), the image created in the image focal plane is planar, and its maximum size is:
\begin{equation}
  h^\text{max}_\text{i}=2f\; ,
  \label{eq:max_size}
\end{equation}
in stark contrast to an ideal lens. This implies that there is no point for the sensor to have a cross-section larger than $2f\times 2f$ (unless one uses external optics to magnify the image to match the sensor size).

\bigskip
The imaging properties of quadratic metalenses found in this section are summarized in Table~\ref{tab:tab}.
\begin{table}
\centering
\begin{tabular}{|c|c|c|}
\hline
Object position & Image coordinates $(x,y,z)$ & Image type \\
%\hline
%\multirow{2}{*}{$0<d_\text{o}<f$} & \emph{Not considered here.} & \multirow{2}{*}{Virtual} \\
%& & \\
\hline
\multirow{2}{*}{$d_\text{o}< f$} & $x, y \in \left[-\infty,\infty\right]$ & \multirow{2}{*}{Hyperboloid (3D virtual image)} \\
& $z\in\left[-\infty,-\frac{d_\text{o}f}{f-d_\text{o}}\right]$ &  \\
\hline
\multirow{2}{*}{$d_\text{o}=f$} & $x, y\in \left[-\infty,\infty\right]$ & \multirow{2}{*}{Infinite} \\
& $z=\infty$ & \\
\hline
\multirow{2}{*}{$d_\text{o}\gtrsim f$} & $x, y \in \left[-f,f\right]$ & \multirow{2}{*}{Ellipsoid (3D image)} \\
& $z\in\left[f,\frac{d_\text{o}f}{d_\text{o}-f}\right]$ &  \\
  \hline
  \multirow{2}{*}{$d_\text{o}\gg f$} & $x, y \in \left[-f,f\right]$ & \multirow{2}{*}{Planar (2D image)}\\
  & $z=f$ &  \\
\hline
\end{tabular}
\caption{Coordinates of image points of an object (second column) and type of obtained image (third column) for different regimes of distances object-lens (first column).
For an object located between the object focal plane and the lens $d_\text{o}<f$, the image is virtual and formed along an hyperboloid; for an object located in the focal plane $d_\text{o}=f$, the image is formed at infinity; for distances larger but comparable to the focal lens $d_\text{o}\gtrsim f$,
  the image is formed along an ellipsoid; for distances much larger than the focal lens $d_\text{o}\gg f$, the image is formed in the image focal plane.}
\label{tab:tab}
\end{table}

%%%%%%%%%%%%%%%%%%%%%%%%%%%%%%%%%%%%%%%%%%%%%%%%%%%%%%%%%%%%%%%%%%%%%%%%%%%%%%%%%%%%%%%%%%%%%%%%%%%%%%%%%%%%%%%%%%%%%%%%%%%%%%%%%%%%%%%%%%%%%%%%%%%%%%%%%%%%%%%%%%%%%%%%
%%%%%%%%%%%%%%%%%%%%%%%%%%%%%%%%%%%%%%%%%%%%%%%%%%%%%%%%%%%%%%%%%%%%%%%%%%%%%%%%%%%%%%%%%%%%%%%%%%%%%%%%%%%%%%%%%%%%%%%%%%%%%%%%%%%%%%%%%%%%%%%%%%%%%%%%%%%%%%%%%%%%%%%%

\section*{Practical implementation and characterization of quadratic metalenses}

\begin{comment}
  The advantage of this geometry is that the coupling with the incident light is quite robust to the incident angle, in comparison to nanoresonators in Huygens' metasurfaces \cite{decker2015high}
for which the coupling with incident light strongly depends on the incident angle \cite{lalanne2017metalenses} that ultimately limits the field-of-view (see Ref.~\cite{engelberg2020near}
where the authors fabricated a quadratic metalens based on silicon Huygens nanoantennas).
Other strategies using photonsieves metasurfaces (e.g., see Ref.~\cite{huang2015ultrahigh}) which are quite robust with incident angle can offer an extremely wide FOV
due to the very small pitch between the ``meta-atoms'' (holes in a metallic film in this case),
but the transmission efficiency is very limited, so that such photonsieve metalenses
offer very low efficiencies (see Ref.~\cite{pu2017nanoapertures} where the author fabricated an extremely large-FOV quadratic metalens using nanoapertures in a gold film, but very low efficiency).
The use of silicon nanopillars was the strategy employed in Ref.~\cite{martins2020metalenses} to design a quadratic metalens working at $\lambda_0=532\,$nm.
\end{comment}

The design of phase-gradient metasurfaces is usually based on the so-called phase-mapping approach, which
consists on mapping a given phase profile %of Eqs.~(\ref{eq:hyp_phase_profile}) and (\ref{eq:par_phase_profile})
with discrete elements, called meta-atoms, which locally impart the desired phase. 
Very importantly, this phase discretization also leads to some limitations of the metalens performances.
In the following, we derive such limitations with regards to the maximum physical diameter and corresponding NA of the lens,
and also give a criterion about the FOV that be can be expected for quadratic metalenses.
We then present the experimental implementation of such lenses to
%explain in Section~\ref{sec:design} how we designed and fabricated our quadratic metalens, and in Section~\ref{sec:exp},
%we show the numerical and experimental characterizations of the designed and fabricated sample, demonstrating
achieve an extremely large-FOV,
%Finally, we end this part with a discussion in Section~\ref{sec:discussion} on
and explain why the actual FOV exceeds by far the one expected from theoretical considerations %FOV given by the criterion,
unveiling the very reason why such quadratic metalenses offer such large FOV.

\subsection*{Design considerations: the effect of the phase discretization}
\label{sec:max_NA}

\subsubsection*{Maximum physical diameter and numerical aperture}

In the phase-mapping approach, the different meta-atoms are enclosed into unit-cells, and a set of unit-cells that samples the phase linearly from 0 to $2\pi$ forms
a super-cell.
Let us denote by $p$ the size (or pitch) of a unit-cell, and let us denote by $N$ the minimum number of unit-cells used to sample the phase from 0 to $2\pi$ in the metalens ($N\geq 2$ to respect Nyquist criterion).
In other words, $N p$ is the length of the smallest super-cell of the metalens. For a metalens with focal length $f$, the maximum physical radius $R^\text{max}$ is reached when the
spatial variation of the phase $\varphi$ is equal to the ratio between $2\pi$ and $N p$:
\begin{equation}
\left|\frac{\partial\varphi}{\partial r}\right|_{R^\text{max}}=\frac{2\pi}{N p}\; .
\end{equation}
Therefore, for a quadratic phase profile [Eq.~(\ref{eq:par_phase_profile})], the maximum physical diameter reads:
\begin{equation}
D^\text{max}=\frac{2f}{n_\text{t}\,\xi}\quad\text{with}\quad\xi\equiv \frac{N p}{\lambda_0}\;.
\label{eq:rmax_par}
\end{equation}
The parameter $\xi$ is an important figure-of-merit to characterize the discretization of the phase profile, and we call it in the following \emph{discretization parameter}.
One can already deduce the lower bound for this parameter: $\xi\geq 1/(n_\text{i}+n_\text{t})$, by recalling that
the maximum reasonable physical diameter of the lens --- when not limited by discretization --- is $D=2f(1+n_\text{i}/n_\text{t})$
[see Eq.~(\ref{eq:phys_diam_ideal})]. In terms of the corresponding maximum physical NA, this leads to:
\begin{equation}
\text{NA}^\text{max}=\frac{n_\text{t}}{\sqrt{1+\left(n_\text{t}\,\xi\right)^2}}\; .
\end{equation}

\subsubsection*{Criterion for field-of-view}

This maximum physical diameter may reduce the FOV of a quadratic metalens in practice.
Indeed, by recalling that the effective working area with radius $R_\text{eff}=f$ is translationally shifted
as the AOI increases (according to Eq.~(\ref{eq:delta_shift})), this area will be cut whenever the AOI exceeds a certain critical angle $\theta_\text{i}^\text{cut}$ satisfying:
\begin{equation}
  \Delta\left(\theta_\text{i}^\text{cut}\right)+R^\text{eff}= R^\text{max}\; .
  \label{eq:fov_eq}
\end{equation}
Thus, we can define a criterion for the minimum FOV of the lens $\text{FOV}\geq 2\,\theta_\text{i}^\text{cut}$, which, by using Eqs.~(\ref{eq:eff_diam_ideal}), (\ref{eq:delta_shift}) and (\ref{eq:rmax_par}) into Eq.~(\ref{eq:fov_eq}), reads:
\begin{equation}
\text{FOV}\geq 2\,\text{arcsin}\left(\frac{1}{n_\text{i}}\frac{1-n_\text{t}\xi}{\xi}\right)\quad \text{with}\quad \frac{1}{n_\text{i}+n_\text{t}} \leq \xi \leq \frac{1}{n_\text{t}}\; .
\label{eq:fov}
\end{equation}
According to this criterion, when the discretization parameter $\xi$ (or the pitch $p$) decreases, the minimum FOV of the lens increases
(see SI Section~4 and Fig.~S5 for a plot of this minimum FOV).
Note that this criterion does not hold if $\xi > 1/n_\text{t}$, since in this case $R^\text{max} < R^\text{eff}$ and the effective working area is already cut at normal incidence.

\subsection*{Experimental realization}
\subsubsection*{Design and fabrication}
\label{sec:design}

\begin{figure}[!h]
\centering\includegraphics[width=13cm]{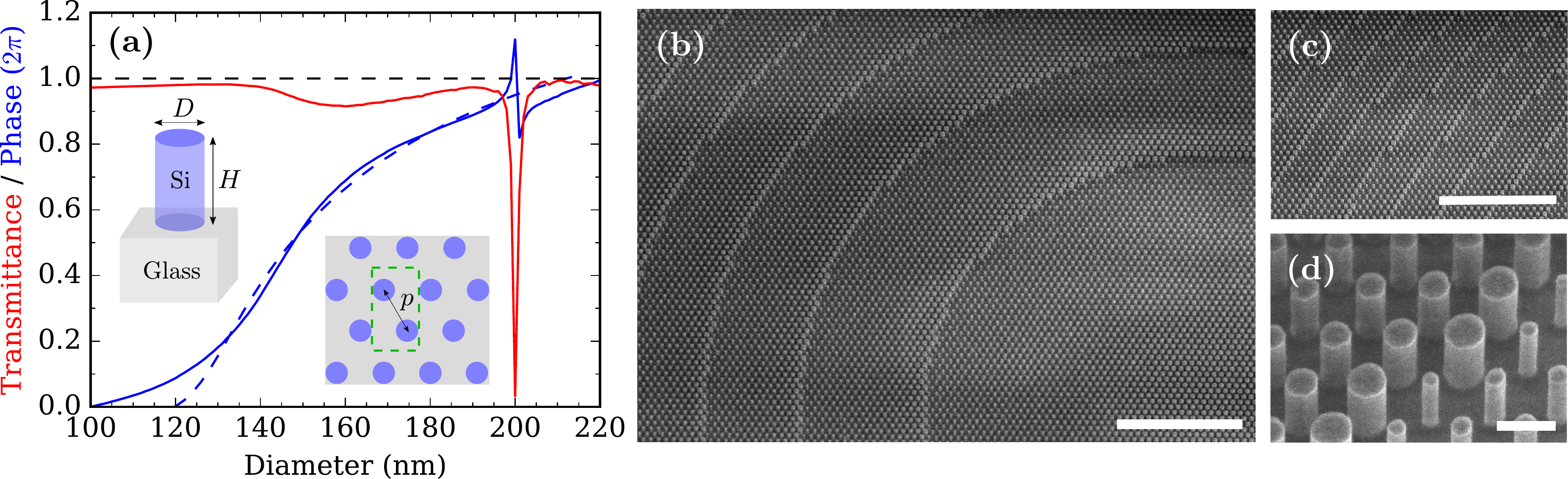}
\caption{(a) Transmission (solid red line) and phase (solid blue line) of a periodic arrangement of identical Si nanopillars as a function of their diameters.
  The left inset represents a single Si nanopillar of height $H$ and diameter $D$ on top of a glass substrate.
  The right inset represents a top-view of the hexagonal lattice with pitch $p$. A unit-cell is also highlighted (dashed green line). 
  The phase accumulated through propagation over a distance $H$ in the case of an infinite circular waveguide is also shown (dashed blue line).
(b-d) SEM images of different parts of the fabricated metalens. The scale bars represent $5\,\mu$m in (b) and (c), and $250\,$nm in (d).} 
\label{fig:fig3}
\end{figure}

We designed and fabricated a quadratic metalens for the working wavelength $\lambda_0=740\,$nm. The meta-atoms used to realize such a quadratic metalens must, ideally, satisfy several conditions:
(i) have a high-refractive index, to ensure a good confinement of the optical fields so that the local phase depends only on the corresponding meta-atom, \emph{i.e.},
minimizing non-local and inter-particle coupling effects; (ii) being lossless, to guarantee high transmission efficiencies;
and (iii) being robust, in terms of transmission efficiency and imparted phase, against the AOI.
As a good meta-atom candidate meeting these requirements, we consider amorphous silicon (aSi) nanopillars with circular cross-sections (see insets in Fig.~\ref{fig:fig3}~(a)).
At $\lambda_0=740\,$nm, aSi presents negligible absorption and a high refractive index ($n=4.02$).
Moreover, such nanopillars act as truncated waveguides which, depending on their diameter, allow guided modes to propagate with different effective refractive indices,
thus imparting different phase-delays to the light \cite{lalanne2017metalenses, arbabi2015dielectric}.

\begin{comment}
As a reference case, for \emph{infinite circular waveguides} (waveguide model) the phase accumulated over a distance $H$ reads:
\begin{equation}
\varphi=\frac{2\pi}{\lambda_0}n_\text{eff}H\; ,
  \label{eq:waveguide_model}
\end{equation}
where $n_\text{eff}$ is the effective refractive index of the guided mode, which can be obtained by solving the transcendental equation of the dispersion relation
(see \emph{e.g.} \cite{paniagua2013enhanced}).
This phase-delay is shown in Fig.~\ref{fig:fig3}~(a) for the \emph{fundamental mode} at $\lambda_0=740\,$nm in aSi circular waveguides and for a distance $H=350\,$nm,
as a function of the waveguide diameter (dashed blue line).
\end{comment}

In Fig.~\ref{fig:fig3}~(a), we simulated using the Finite-Difference Time-Domain (FDTD) method (see Methods section) the phase imparted by a periodic array of aSi nanopillars
with height $H=350\,$nm for a normally incident wave, as a function of the nanopillar diameters (blue line). %, which matches quite well with the waveguide model.
The transmittance of the array is also shown (red line). One can see that by varying the diameter, one can span the whole $2\pi$-phase space while maintaining a good transmission efficiency.
We also studied the influence of the AOI of the incident wave, and found that phases and transmittances are quite robust (see SI Section~5 and Fig.~S6).
For the simulations, we considered a hexagonal lattice with pitch $p=300\,$nm (distance between the centers of the nearest pillars), to ensure that near-field interactions between nanopillars are negligible and that the behavior of a single nanopillar is mostly independent from its neighbours.
This guarantees the validity of the local phase-mapping approach, which assumes that the meta-atoms are surrounded by similar meta-atoms (local periodicity approach), and allows
to extrapolate these simulated transmittances and phases to the metalens design. 

For the design of the quadratic metalens, we therefore choose nanopillars with height $H=350\,$nm and
different diameters in the range $D=[100\,\text{nm},220\,\text{nm}]$ in order to map the $2\pi$-wrapped phase profile given in Eq.~(\ref{eq:par_phase_profile}). We purposely avoid the portion $D=[195\,\text{nm},205\,\text{nm}]$,
which corresponds to a sharp Fabry-Perot resonance leading to low transmission values (see Fig.~\ref{fig:fig3}~(a)).
With this library of nanopillars, we fabricated a quadratic metalens with a diameter $D=500\,\mu$m and a discretization parameter $\xi=0.81$, corresponding to a smallest super-cells containing $N=2$ nanopillars
and leading to a focal length $f=203\,\mu$m (see Eq.~(\ref{eq:rmax_par})).
The manufacturing of the metalens is done using electron beam lithography,
and deep reactive ion etching, using a pseudo-Bosch process \cite{hung2010fabrication}, providing smooth sidewalls with minimized tappering (more details in the Methods section).
In Fig.~\ref{fig:fig3}~(b-d), we show representative scanning electron microscope (SEM) images of different parts of the metalens taken with different magnifications.

\subsubsection*{Optical characterization}
\label{sec:exp}

\begin{figure}[!ht]
	\centering\includegraphics[width=\linewidth]{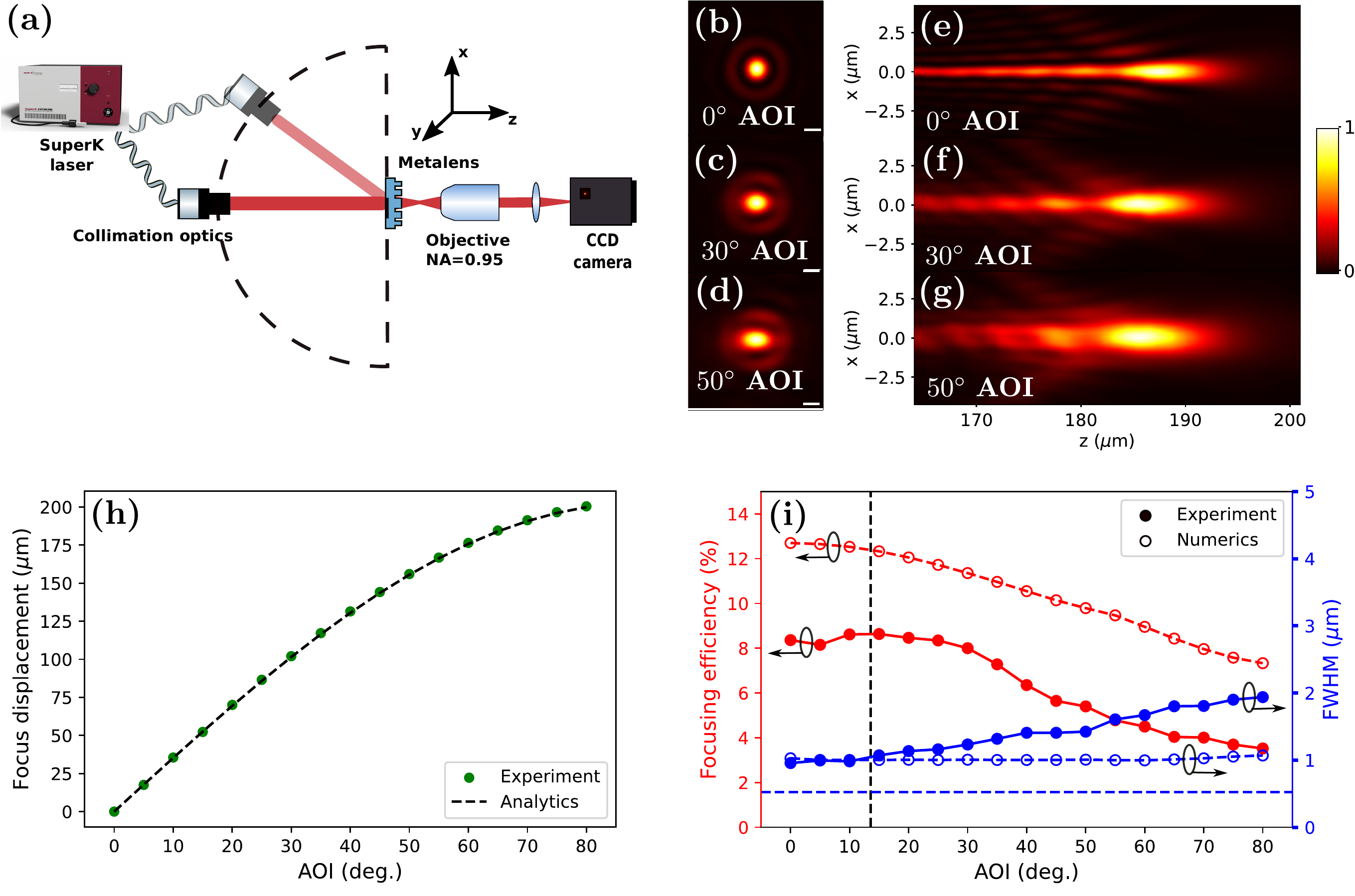}
	\caption{Metalens optical characterization. (a) Schematic of the optical setup for characterization.
          (b), (c) and (d) Normalized PSF images for $0^\circ{}$, $30^\circ{}$ and $50^\circ{}$ AOI. The scale bars correspond to $1\,\mu$m.
          (e), (f) and (g) Normalized intensity distributions measured in the $x$-$z$ plane for $0^\circ{}$, $30^\circ{}$ and $50^\circ{}$ AOI.
          (h) Lateral displacement of the PSF in the $x$-direction (green dots) as a function of the AOI (green dots). The theoretical prediction
          [see Eq.~(\ref{eq:delta_shift})] is also shown (black dashed line).
          (i) Focusing efficiency (red dots, left axis) and FWHM of the PSF (blue dots, right axis) as a function of the AOI, together with
          the corresponding numerical simulations (open red and blue circles, respectively). The lines are guide-to-the-eye.
          The FWHM corresponding to a diffraction-limited lens with NA$\simeq 0.71$ [see Eq.~(\ref{eq:eff_NA})] is also shown for reference (blue dashed line).
          The critical angle $\theta^\text{cut}_\text{i}$ [see Eqs.~(\ref{eq:fov_eq}) and (\ref{eq:fov})] is indicated by the vertical black dashed line.
}
	\label{fig:fig4}
\end{figure}

We explore the performance of our fabricated quadratic metalens by a comprehensive analysis of the experimental point spread function (PSF),
an observable that fully describes the imaging properties of an optical system. The optical setup used for this analysis is depicted in Fig.~\ref{fig:fig4}~(a) (more details in the Methods section).
We first show the normalized PSF images produced by a laser beam illumination for $0^\circ{}$, $30^\circ{}$ and $50^\circ{}$ AOI, in Figs.~\ref{fig:fig4}~(b), (c) and (d), respectively.
One can see good quality focal spots even for large angles, though widening of the PSF across the horizontal direction ($x$-axis) is apparent at $50^\circ{}$ AOI.
We also show the light intensity distribution across the OA for $0^\circ{}$, $30^\circ{}$ and $50^\circ{}$ in Fig.~\ref{fig:fig4}~(e), (f) and (g), respectively.
This is done by scanning the metalens position along the $z$-axis and recording multiple PSF that are then stitched to reconstruct the corresponding $x$-$z$ intensity maps for different AOI.
One can see that the position of the maximum intensity along the optical axis remains largely unaltered, even for larger angles.

To provide a more quantitative picture of the metalens performance, we measured the PSF lateral displacement for an extended range of AOI up to $80^\circ{}$, as shown in Fig.~\ref{fig:fig4}~(h).
One can see a perfect agreement between the experimental measurements (green dots) and the analytical formula given by Eq.~(\ref{eq:delta_shift}) (black dashed line).
We also measured the focusing efficiency (red dots), and the FWHM of the PSF (blue dots) as a function of the AOI, as shown in Fig.~\ref{fig:fig4}~(i).
As introduced in Section~\ref{sec:focusing}, the former is defined as the power
within the focal spot (integrated over the second Airy disk) divided by the power passing through the effective working area of the lens, with diameter $D^\text{eff}=2f$.
In the plot, one can see that the efficiency remains virtually unchanged up to $30^\circ{}$ and slowly decreases for higher angles, due to the fact that the effective working area
is getting increasingly cropped by the physical size of the lens as the AOI increases.
We compare these results with Fourier optics simulations of a quadratic metalens with the same lens size $D=500\,$nm
and focal length $f=203\,\mu$m, and discretized transmissions and phases with pitch $p=300\,$nm, whose values are taken from Fig.~\ref{fig:fig3}~(a).
While the general trend in the experiment is similar to the numerical simulations (open red circles),
the absolute values of the efficiency are smaller than the theoretical ones, which we attribute to the fabrication imperfections.
As far as the FWHM is concerned, we measured a FWHM of about $1\,\mu$m at normal incidence, perfectly correlating with the numerical simulations (open blue circles). 
With increasing AOI, however, one can see a slight increase of the FWHM, both experimentally and theoretically, but less pronounced in theory.
The increase of the FWHM of the PSF can be explained by the fact that coma aberrations are introduced when the effective working area is getting cropped, thus affecting the quality of the focal spot,
in accordance with our aberration analysis of quadratic metalenses (see SI Section~1).
The discrepancy between the experiment and theory might come from the AOI dependence of the nanopillar transmissions and phases (see SI Section~5), as well as the vectorial nature of light, that are not taken into account
in these numerical simulations.
%in accordance to the predicted transmission drop of the meta-atoms for large AOI (see \textcolor{red}{SI Section~3 and Fig.~S5}).

\subsubsection*{A note on the criterion for field-of-view}
\label{sec:discussion}

The actual FOV of our quadratic metalens exceeds by far the criterion given by Eq.~(\ref{eq:fov}). According to this criterion, with the current discretization parameter $\xi=0.81$,
the minimum FOV after which the effective working area is getting cut by the physical size of the lens is $\text{FOV}\geq 27^\circ{}$ (denoted by the vertical black dashed line in Fig.~\ref{fig:fig4}~(i)).
As can be seen, however, the experiment indicates that both the efficiency and the quality of the focus are still reasonable even after the minimum AOI has doubled.
This is actually not surprising since, as explained in Section~\ref{sec:focusing}, the quadratic metalens produces a PSF that is not diffraction-limited, due to its intrinsic spherical aberrations.
For our design, the PSF has a FWHM of $1\,\mu$m, at least at normal incidence, which corresponds to the PSF produced by a diffraction-limited system with an $\text{NA}\simeq 0.37$,
which is much smaller
than the effective NA of quadratic metalenses: $\text{NA}^\text{eff}\simeq 0.71$ (see Eq.~(\ref{eq:eff_NA})).
In other words, the area that contributes to focusing is much smaller than the effective working area. 
Also, this is why the focal spot
is not affected even when the effective working area starts to be cropped by the metalens, which was the criterion used to derive Eq.~(\ref{eq:fov}).
Therefore, as we can see, quadratic metalenses might have a FOV exceeding this criterion, provided that their physical diameter $D$ is larger than the effective diameter $D^\text{eff}=2f$.

%%%%%%%%%%%%%%%%%%%%%%%%%%%%%%%%%%%%%%%%%%%%%%%%%%%%%%%%%%%%%%%%%%%%%%%%%%%%%%%%%%%%%%%%%%%%%%%%%%%%%%%%%%%%%%%%%%%%%%%%%%%%%%%%%%%%%%%%%%%%%%%%%%%%%%%%%%%%%%%%%%%%%%%%
%%%%%%%%%%%%%%%%%%%%%%%%%%%%%%%%%%%%%%%%%%%%%%%%%%%%%%%%%%%%%%%%%%%%%%%%%%%%%%%%%%%%%%%%%%%%%%%%%%%%%%%%%%%%%%%%%%%%%%%%%%%%%%%%%%%%%%%%%%%%%%%%%%%%%%%%%%%%%%%%%%%%%%%%

\section*{Imaging with a quadratic metalens}

In this Section, we use the fabricated quadratic metalens in different imaging configurations, unveiling the fundamental limitation that constrains their performance, despite
the extremely large FOV: the barrel distortion or fish-eye effect.
%In particular, taking into account this effect, we derive analytical formulas to show the existing trade-off
%between the maximum FOV of an object that can be imaged and the optical spatial resolution of the metalens.
We first discuss this effect on the imaging of a ruler, and illustrate the existing trade-off
between the maximum FOV of an object that can be imaged and the optical spatial resolution of the metalens.
With these limitations in mind, we then apply our metalens to the problem of fingerprint capturing, which requires quite an extreme imaging configuration in terms of distance and FOV.

\subsection*{Barrel distortion}
\label{sec:barrel}

\begin{figure}[!ht]
\centering\includegraphics[width=13cm]{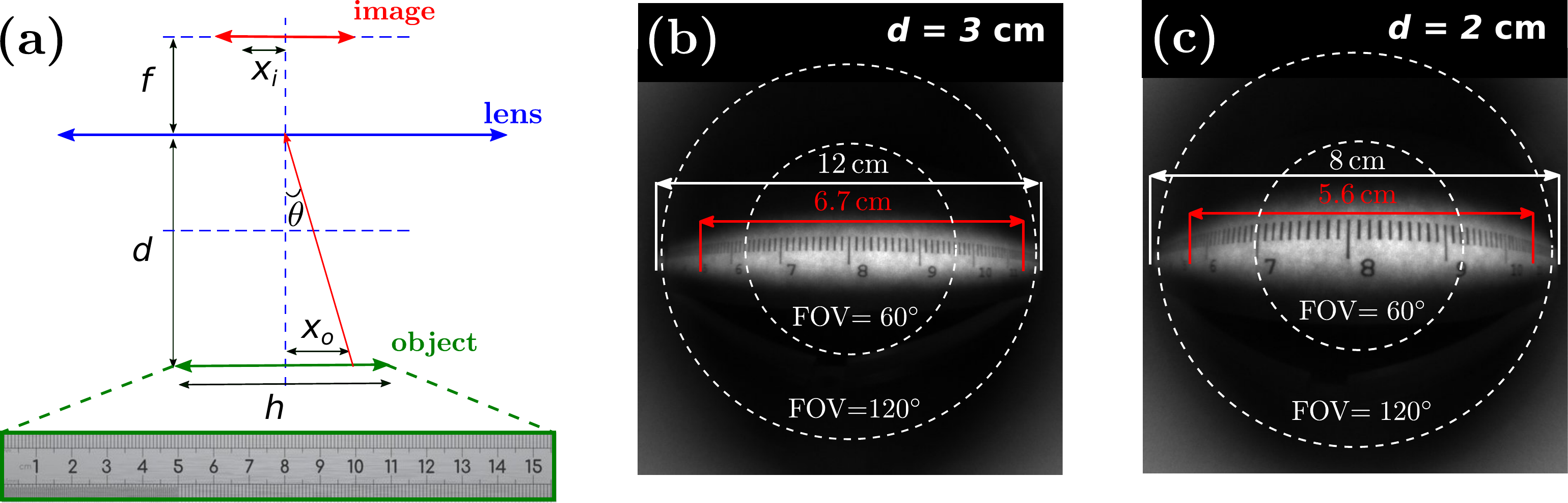}
\caption{Illustration of the barrel distortion effect on the imaging of a ruler. (a) Parameters of the imaging configuration: $f$ is the focal length of the metalens; $d$ is the distance ruler-metalens; $x_\text{o}$ denotes the distance from the OA of a given object point to be imaged; $x_\text{i}$ is the distance from the OA of the corresponding image point; $h$ is the size of the ruler's part being imaged. (b), (c) Images of a ruler taken with our quadratic metalens (focal lens $f=203\,\mu$m) in the following configurations:
  $d=3\,$cm and $d=2\,$cm, respectively. The white arrows correspond to parts of the ruler that are being imaged but not entirely resolved, and the red arrows correspond to the parts of the ruler for which the ruler graduations are fully resolved, and their lengths $h$ have been calculated from Eqs. (\ref{eq:FOV_vs_res}) and (\ref{eq:FOV_exp}) by taking the object feature size $\Delta x_\text{o}\sim 1\,$mm and the image spatial resolution $\Delta x_\text{i}\sim 2\,\mu$m.}
\label{fig:barrel_distortion}
\end{figure}

We start by imaging a ruler located at a distance $d$ from the metalens that verifies $d\gg f$,
in order to be in the conditions of imaging a planar object into a planar image (see Table~\ref{tab:tab}).
The situation is illustrated in Fig.~\ref{fig:barrel_distortion}~(a) (not to scale).
We consider two configurations: $d=3\,$cm and $d=2\,$cm.
The resulting images obtained with our fabricated metalens are shown in Figs.~\ref{fig:barrel_distortion}~(b) and (c), respectively.
One can see that parts of the ruler of sizes approximately equal to $h\sim 12\,$cm and $h\sim 8\,$cm are being imaged, respectively.
This corresponds to an actual $\text{FOV} > 120^\circ{}$ in both configurations, which is estimated using the following formula (valid when $d\gg f$):
\begin{equation}
  \text{FOV}\simeq 2\,\text{arctan}\left(\frac{h}{2d}\right) \; .
  \label{eq:FOV_exp}
\end{equation}

In Figs.~\ref{fig:barrel_distortion}~(b) and (c), the barrel distortion is visible, which is characteristic of wide-FOV lenses (\emph{e.g.} fisheye lenses) \cite{hughes2008review,kim2015fisheye,sahin2018distortion}.
This follows from the fact that equally spaced object points produce image points that get more compressed in the edges of the image. 
The barrel distortion in itself is not a problem for imaging since it can be compensated with a post-processing treatment (see next Section). However, as the distance of the object points $x_\text{o}$ increases, or in other words, as the FOV of the object increases (according to Eq.~(\ref{eq:FOV_exp}) with $h=2x_\text{o}$), the corresponding image points
become increasingly compressed (see SI Section~6 and Fig.~S7).
Subsequently, the image spatial resolution required to resolve these points will increase. In practice, the image spatial resolution  corresponds, depending on which one is the limiting factor, to either the detector spatial resolution (related to the pixel pitch), or to the optical spatial resolution of the lens (proportional to the FWHM of the PSF), which is its ability to resolve details.
In order to be more quantitative, we derived explicitly (see SI Section~6 and Fig.~S8) the maximum FOV of an object over which features with a given minimum size $\Delta x_\text{o}$ can be resolved with a quadratic metalens, and for a given image spatial resolution $\Delta x_\text{i}$:
\begin{equation}
  \text{FOV}=2\,\text{arctan}\left(\sqrt{\left(\frac{f}{d}\frac{\Delta x_\text{o}}{\Delta x_\text{i}}\right)^{\frac{2}{3}}-1}\right)\; .
  \label{eq:FOV_vs_res}
\end{equation} 

%In Fig.~\ref{fig:barrel_distortion}~(e), we plot this maximum FOV as a function of the relative image spatial resolution $\Delta x_\text{o}/\Delta x_\text{i}$. One can see that the maximum FOV increases as the
%relative image spatial resolution $\Delta x_\text{o}/\Delta x_\text{i}$ increases.

In view of this, we analyse the previous ruler imaging experiment.
%By inverting Eq.~(\ref{eq:criteriabis}), we find the threshold $x^\text{thres}_\text{o}$ above which, for a given image resolution $1/\Delta x_\text{i}$,
%object features larger than $\Delta x_\text{o}$ cannot be resolved:
%\begin{equation}
%x^\text{thres}_\text{o}=d\sqrt{\left(\frac{f}{d}\frac{\Delta x_\text{o}}{\Delta x_\text{i}}\right)^{2/3}-1}\;.
%   \label{eq:criteria2}
%\end{equation}
We take the minimum object feature size to be $\Delta x_\text{o}\sim 1\,$mm.
Since in our imaging configuration, we are not limited by the detector spatial resolution
(for the details of the imaging experiment, see SI Section~7 and Fig.~S9), we take for the image spatial resolution the worse FWHM of the PSF, that we measured to be $\Delta x_\text{i}\equiv \text{FWHM}\sim 2\,\mu$m
(obtained for an AOI of $80^\circ{}$, as seen in Fig.~\ref{fig:fig4}~(b)).
The application of Eq.~(\ref{eq:FOV_vs_res}) gives the maximum FOV of the ruler over which such $1\,$mm features can be resolved. They turn out to be $\text{FOV}\simeq 96^\circ$ in the configuration with $d=3\,$cm and
$\text{FOV}\simeq 109^\circ$ in the configuration with $d=2\,$cm, corresponding to portions of the ruler with sizes equal to $h=6.7\,$cm and $h=5.6\,$cm, respectively (according to Eq.~(\ref{eq:FOV_exp})).
%(these cases are shown in Fig.~\ref{fig:barrel_distortion}~\textcolor{red}{(e)} by the left and right vertical dashed lines, respectively), which
%require $\text{FOV}<100^\circ{}$ and $\text{FOV}<110^\circ{}$, respectively.
One can indeed check with naked eye on the images of Figs.~\ref{fig:barrel_distortion}~(b) and (c) that the ruler graduations can only be
properly resolved for those parts of the rulers with sizes
$h<7\,$cm and $h<6\,$cm, respectively.

This example illustrates that it is the barrel distortion that limits the FOV of the lens in an imaging configuration, which, in turn, is ultimately constrained by the image spatial resolution required
to resolve details of an object.
In the example of the ruler located at two different distances $d=3\,$cm and $d=2\,$cm from our fabricated lens, while portions of the ruler with sizes $h\sim 12\,$cm and $h\sim 8\,$cm, respectively,
can be imaged thanks to the large FOV of the metalens,
its optical spatial resolution only allows to resolve parts with sizes $h=6.7\,$cm and $h=5.6\,$cm, respectively, with the desire $1\,$mm spatial resolution.

\subsection*{A practical case: fingerprint imaging}
\label{sec:example}

\begin{figure}[ht]
	\centering\includegraphics[width=5.7cm]{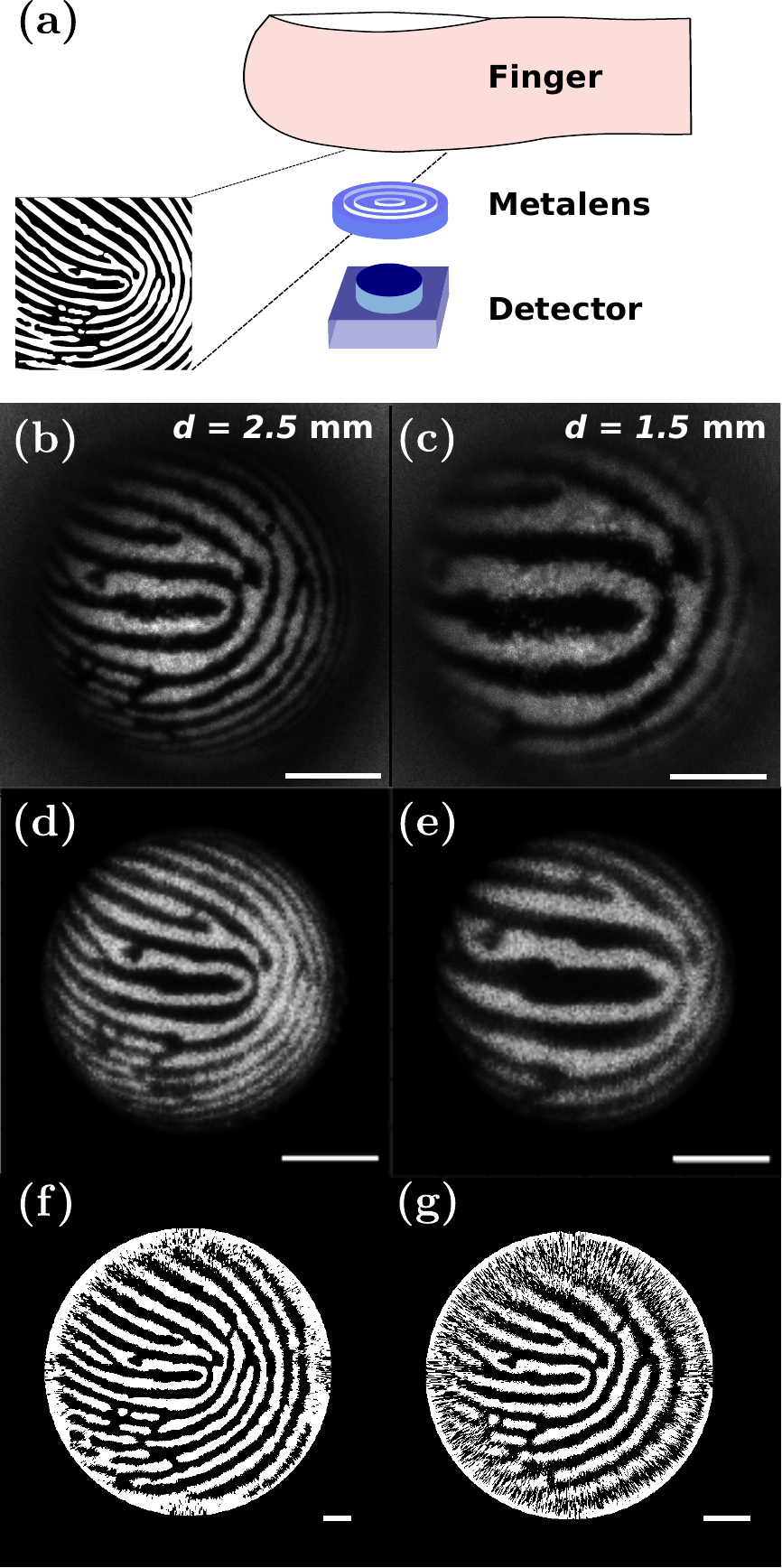}
	\caption{Fingerprint imaging. (a) Artistic picture of the fingerprint capturing device and original image of the printed fingerprint (not to scale). (b), (c) Experimental images
          of the fingerprint taken with our fabricated metalens at distances $d=2.5\,$mm and $d=1.5\,$mm, respectively. (d), (e) Corresponding simulated images.
          (f), (g) Post-processed
          experimental images (b) and (c), respectively. The scale bars correspond to $100\,\mu$m.}
	\label{fig:fig6}
\end{figure}

In this section, we investigate more extreme imaging configurations (i.e. shorter distances $d$) in the particular problem of fingerprint capturing,
to demonstrate the imaging potential of quadratic metalenses. Indeed, the subwavelength scale of the metalens opens up new possibilities to realize compact optical detection systems possessing large FOV (see the artistic picture on Fig.~\ref{fig:fig6}~(a), left panel),
which remains elusive for conventional refractive optics, as it typically involves complex optical trains.
In order to scale down the device thickness, the object-lens distance $d$ should be reduced as much as possible, which in turn requires higher FOV for a given fingerprint size.
Taking into account the barrel distortion constraint, the object-lens distance obviously possesses a lower limit, below which one can no longer resolve the object features.
Here, we examine this limit with our fabricated metalens by imaging a $5\,\text{mm}\times 5\,$mm fingerprint featuring details $\sim 100\,\mu$m.  

The fingerprint (see Fig.~\ref{fig:fig6}~(a)) was printed on a white paper and illuminated by a laser in a transmission configuration.
Because of the small focal lens of our metalens and the bulky detector that we used, we had to transfer the image produced by the metalens to a CCD camera using a free-space microscope setup with 5X magnification
(for the details of the imaging experiment, see Methods and SI Section~7 and Fig.~S9).
In applications, however, in order to achieve a compact system, it is not desirable to use a microscope and one would place a small detector directly in the image focal plane of the metalens instead.
Figs.~\ref{fig:fig6}~(b) and (c) show the experimental images
for two configurations: $d=2.5\,$mm and $d=1.5\,$mm, respectively. We also provide the corresponding simulated images in Figs.~\ref{fig:fig6}~(d) and (e). The imaging simulation was done using Fourier optics and assuming a perfect quadratic phase profile for the metalens
(see Methods). Moreover, in order to mimic the incoherent imaging, the fingerprint amplitude mask was multiplied with a random phase mask and averaged over realizations.
%One can see that despite a good correspondence, the simulated images have slightly better FOV. We attribute this to the reduction of the focusing efficiency and the PSF broadening at larger AOI.
One can see a good correspondence between the simulated images and the experimental ones.
As expected, all the images clearly present the barrel distortion effects. However, by a proper post-processing (see Methods), it is possible to retrieve corrected images from the experimental ones,
as shown in Figs.~\ref{fig:fig6}~(f) and (g). One can see that the whole $5\,\text{mm}\times5\,$mm fingerprint area becomes visible at $d=2.5\,$mm, while
for the case of $d=1.5\,$mm, the edges are blurred by the barrel distortion and the corresponding information is lost, and thus only a fingerprint area of about $3\,\text{mm}\times 3\,$mm is detected.
In the former case, taking into account the distance between the lens and the image plane,
the total thickness of the device is still only $\sim2.7\,$mm, making for a very compact system for fingerprint detection.

In principle, of course, the object-lens distance $d$ could be further decreased in an attempt to make an even more compact device. This, however, comes at the cost of imaging a smaller object size.
To do so, the focal length should be decreased accordingly, in order to remain in the regime $d\gg f$ of imaging a planar object into a planar image.
To illustrate this scenario, we fabricated another metalens with smaller focal length while keeping the same diameter/focal length ratio ($f=100\,\mu$m, $D=246\,\mu$m).
Using this lens, we imaged the fingerprint at the smallest possible distance $d=800\,\mu$m, which corresponds to the
thickness of the substrate (see Methods, SI Section~7 and Fig.~S9).

\section*{Conclusion}

In this work, we have shown that in terms of imaging, a quadratic metalens images a \emph{planar} object along an \emph{ellipsoid} surface in general, and that in order to produce a \emph{planar} image from a \emph{planar} object,
the object must be located at a distance $d$ from the metalens that satisfies $d\gg f$, $f$ being the focal length of the metalens. In this case, the image has a maximum cross-section of $2f\times 2f$, which also sets the maximum size of the sensor (if no magnification system is used).
Therefore, one must keep in mind that while quadratic metalenses can favor miniaturization (and also cost matters since smaller sensor sizes are cheaper), they can also decrease the resolution, so that only for applications which do not require large sampling, like \emph{e.g.} in the imaging of fingerprints, one can benefit from them. On the other hand, the relatively large depth-of-focus of quadratic metalenses could be beneficial for certain applications, such as in biological imaging, where the depth-of-focus matters to obtain morphological information about tissues for example \cite{zhang2021metasurfaces}. We should also point out here that the image contrast can be improved by placing an aperture stop before the quadratic metalens (see \emph{e.g.}, Ref.~\cite{engelberg2020good}, and a discussion of such configuration based on a modulation transfer function analysis in SI Section~8 and Fig.~S10).

\begin{comment}
in terms of focusing, such metalenses have a maximum effective working area corresponding to a numerical aperture of $\text{NA}\simeq 0.71$ in air
(independent of the focal length
and diameter of the lens), that produces a focal spot that moves laterally in the image focal plane as the AOI increases.
The focal spot is elongated in the longitudinal direction, which is characteristic of spherical aberrations, intrinsic to
quadratic metalenses.
Consequently, we found that the focal spot has a FWHM that is about two times larger than the one
produced with a diffraction-limited lens of $\text{NA}\simeq 0.71$
(for our designed and fabricated metalens working at the wavelength $\lambda_0=740\,$nm and focal length $f=203\,\mu$m, we found a FWHM of about $1\,\mu$m, both in theory and experimentally).
Moreover, we found that the focusing efficiency defined as the ratio between the energy in the focal spot and the energy passing through the effective working area
is rather low (about $14\%$ in theory and $8\%$ experimentally), which is also an intrinsic limitation of this phase profile.
On the other hand, the FOV of quadratic lenses can, in theory, reach the maximum value of $\text{FOV}=180^\circ{}$.
Based on the particular design used, our fabricated metalens has a $\text{FOV}>120^\circ{}$, being limited due to the discretization of the phase in the phase-mapping approach used.
\end{comment}

We also confirmed that quadratic metalenses are excellent flat optical solutions for wide FOV imaging. Indeed, we designed and fabricated a sample working in the near IR, which, despite a rather low focusing efficiency (about $14\%$ in theory and $8\%$ experimentally) which is peculiar to this type of phase profile, allowed us to image a ruler over a FOV of about $100^\circ{}$.
In addition, we highlighted that it is the optical resolution of the metalens and/or the detector resolution that prevents the FOV to be larger, which stems from the intrinsic barrel distortion, or fisheye effect.

As an example and proof-of-concept of an imaging configuration in which the benefits of these metalenses are exploited, we show, using our fabricated sample, the imaging of a fingerprint of size $\sim 5\,$mm for
a distance object-lens as small as $d=2.5\,$mm,
with which we are able to resolve features with sizes of the order of $100\,\mu$m, resulting in a total device thickness of $\sim 2.7\,$mm.
This demonstrates the ability of quadratic metalenses for imaging in extreme configurations of large objects at relatively short distances.

\newpage
\section*{Methods}

\paragraph{Fourier optics calculations:}
The Fourier optics calculations used in this work are based on the \emph{plane-wave spectrum method} (see \emph{e.g.} \cite{goodman2005introduction}, Chapter 3). The so-called plane-wave spectrum or angular spectrum representation is obtained by applying scalar wave approximation and solving the Helmholtz equation.
The method is implemented using transfer function representation in order to exploit discrete Fourier transforms and multiplicative relationships in frequency domain for the calculation of
field propagation. Parabolic lens profiles, defined by wavelength of operation, focal length and lens diameter according to Eq.~(\ref{eq:par_phase_profile}),
are numerically characterized using this plane wave decomposition technique. In oblique angle of incidence calculations, complex fields of an incident plane wave are superimposed with
fields corresponding to the parabolic phase profile. Subsequent propagation of the fields comprises Fourier transform, multiplication by transfer function and inverse Fourier transform.
Transmissive areas of the parabolic lens are visualized by applying a low-pass filter to reduce spatial frequencies to propagating components.
Calculations illustrating light propagation in the $z$-direction are reduced to two dimensions ($x$ and $z$) in order to reduce computational cost.
Point-like sources with a high angular spread are realized using a Gaussian intensity profile with a sub-wavelength FWHM of $500 \,$nm.
Angle dependent efficiency of the lens as well as FWHM and spatial position of foci are calculated in the focal plane given by $z = 194 \,\mu$m.
Efficiency is calculated as the ratio between the power incident on the total lens area and the power in a circle up to the second Airy minima.
For the imaging configurations, a single snapshot of incoherent light propagation is modeled by assigning a pseudo-random phase to each pixel in the object plane and recording propagated fields in the image plane.
The full image is reconstructed in a Monte Carlo like simulation in which multiple snapshots using different random phase masks are averaged.
It is important to note that these Fourier optics simulations require a fine sampling of the object and image planes.
In all calculations, we used the design wavelength of $740 \,$nm, a spatial resolution of at least $150 \,$nm and lateral calculation domain size of at least $1\,$mm $\times$ $1\,$mm.

\paragraph{FDTD simulations:}
We use Lumerical FDTD software for the FDTD simulations of the transmission efficiencies and phase-delays of periodic arrays of silicon nanopillars with identical diameters.
We consider an hexagonal lattice with lattice constant $p=300\,$nm (center-to-center distance between closest nanopillars), and we use periodic boundary conditions at normal incidence or Bloch boundary conditions at oblique incidences.
The pillars are lying on a glass substrate of refractive index $n=1.46$.
We simulate a monochromatic plane wave with wavelength $\lambda_0=740\,$nm and amplitude equal to unity, coming from the glass substrate (incident medium $n_\text{i}=1.46$), and
record the transmission efficiencies and phases after the propagation through the nanopillars in air (transmission medium $n_\text{t}=1$), and we sweep over the nanopillar diameters and/or AOI.

\paragraph{Fabrication:}
To fabricate our samples, we first deposit a $350\,$nm thick layer of amorphous silicon on $800\,\mu$m thick fused silica substrates using chemical vapor deposition technique
(model Producer SE, Applied Materials) followed by a $30\,$nm thick layer of chromium (Cr) by evaporation (Evovac, Angstrom Engineering)
to act later as a hard mask for silicon etching. For the mask nanopatterning, a thin layer of hydrogen silsesquioxane (HSQ, Dow Corning, XR-1541-002)
is formed by spin-coating and baking. Electron-beam lithography (Elionix, 100 kV) is used to expose the negative resist, and the development in Tetramethylammonium hydroxide solution
(TMAH, 25\%) reveals the metalens design. The pattern is then transferred from the resist to the Cr hard mask using ICP-RIE (Plasmalab System 100, Oxford Instruments)
with chlorine chemistry. The choice of using Cr as mask for silicon etching is to get rid of the remaining HSQ. The resist thickness ($50\,$nm) is then chosen to no longer survive the silicon etching.
%For the silicon etching, Bosch process is commonly used, which consists of a first isotropic etching step using SF6 gas and the second step of a passivation layer deposition under C4F8 gas.
%This process allows a deep anisotropic etching but suffers from the well-known scalloping effect not suitable for nanostructure fabrication.
%However, to obtain an anisotropic etching with smooth sidewalls, the samples are etched using both etch/passivation chemistries simultaneity, also called pseudo-Bosch process,
%which after optimization of the chamber parameters (SPTS Technologies) allows precise control of the sidewall angle of the nanopillars.
To obtain anisotropic etching with smooth sidewalls, the samples are etched using the so-called pseudo-Bosch process, in which etching (SF$_{6}$ gas) and passivation (C$_{4}$F$_{8}$ gas) chemistries are used simutaneously.
Optimization of the chamber (SPTS Technologies) parameters allows precise control of the sidewall angle of the nanopillars.
The chromium mask is then removed by Cr etchant (Sigma-Aldrich).

\paragraph{Optical measurements:}
We performed the PSF optical characterization using a tunable fiber laser centered at 740 nm wavelength (SuperK NKT Photonics equipped with a tunable single line filter SuperK Select).
The fiber output together with collimation optics were placed on a rotation stage (see Fig.~\ref{fig:fig4}~(a)). A linear polarizer was utilized to produce a TM-polarized laser beam.
The PSF of the metalens was imaged by a free-space 83$\times$ microscope setup (100$\times$ Olympus plan apo infinity-corrected objective, NA=0.95, a tube lens with 150 mm focal length and CS895MU Thorlabs
CMOS camera). The focusing efficiency was calculated by integration of the PSF intensity over a circular aperture with a constant radius, which corresponds to the second Airy minimum position for
AOI = $0^{\circ}$. Anti-reflection coated achromatic doublet (Thorlabs AC254-030-B-ML) was utilized as a reference lens.

For imaging the ruler and fingerprint, we used a free-space 5$\times$ microscope setup
depicted on SI Fig.~S9. The objects were illuminated by the laser in transmission configuration. To produce a line pattern for the ruler illumination,
we used a Thorlabs $0.4^{\circ}\times 100^{\circ}$ line engineered diffuser. The post-processing of the fingerprint images was performed by the following algorithm:\\
1. The radius $R$ and the center of distortion were extracted from the raw image. The maximum radius of new undistorted image was calculated as
$R_\text{max}=f\,\text{tan}(\text{arcsin}(R/f))\; ;$ \\
2. The undistorted image pixels $(i,j)$ were filled in by the intensity values extracted from the raw image $(m,k)$. $(m,k)$ pixels were identified by the following formula:\\
$m,k=\frac{i-R_\text{max}}{\sqrt{(i-R_\text{max})^{2}+(j-R_\text{max})^{2}}}\,f\, \text{sin}\left\{\text{arctan}\left(\frac{\sqrt{(i-R_\text{max})^{2}+(j-R_\text{max})^{2}}}{f}\right)\right\}+R-1\; ;$\\
3. The new undistorted image was binarized using an adaptive thresholding technique suitable for a non-uniform intensity distribution.

%\vskip20mm
\section*{Authors' contributions:}
E.L. designed the metalens, made the FDTD simulations, and carried out the theoretical analysis of the imaging properties.
T.W.W.M. implemented the Fourier optics code, made all the Fourier optics calculations (unless otherwise stated), and took part in the theoretical analysis of the imaging properties.
D.E. optimized the fabrication process and fabricated all the samples.
A.V.B. performed the optical characterization, the imaging experiments, the Fourier optics calculations for the fingerprint and the calculations of the modulation transfer function.
E.K. performed the aberration analysis of the quadratic metalenses based on Zernike polynomials.
S.L. supervised the fabrication. R.P.-D. implemented the ray-tracing code.
R.P.-D. and A.I.K. conceived the idea and supervised the work.
E.L., T.W.W.M., D.E. and A.V.B. contributed equally to this work and wrote the manuscript. All the authors read and reviewed the manuscript.

\begin{acknowledgement}

The authors thank Steven Lee Hou Jang for depositing the amorphous silicon on a $12"$ fused silica wafer, and Moitra Parikshit for providing accurate values of the refractive index and extinction coefficient of amorphous silicon via ellipsometry measurements.
The authors acknowledge financial support from National Research Foundation of Singapore (grant NRF-NRFI2017-01),
IET A F Harvey Engineering Research Prize 2016, A*STAR SERC Pharos program (grant 152 73 00025) (Singapore),
and AME Programmatic Grant A18A7b0058 (Singapore).

\end{acknowledgement}

\begin{suppinfo}

Analysis of the monochromatic aberrations of quadratic metalenses: Zernike decomposition of the wavefront and ray-tracing pictures for different AOI. Fourier optics calculations of the focal spot position, focusing efficiency and FWHM as a function of the AOI. Imaging theoretical analysis: derivation of the ellipse and hyperbola equations in the paraxial approximation. Plot of the minimum metalens FOV as a function of the discretization parameter/unit-cell pitch. Additional information on the transmission and phase of the meta-atoms for different AOI and for TM/TE polarizations: transmission and phase maps for periodic array of unit-cells, diffraction efficiencies into the blazed order for periodic array of super-cells. Derivation and plot of the equation for the maximum object FOV in terms of the image spatial resolution. Additional information about the imaging experiment of the ruler and the fingerprint. Analysis of the monochromatic modulation transfer function.

\end{suppinfo}

\bibliography{biblio}

\end{document}

\noindent
Number of pages: 16\\
Number of figures: 10\\
Number of tables: 1

\section{Aberration analysis}

Zernike polynomials compose a complete orthogonal set on a unit circular aperture to decompose the wavefront of light.
Such a wavefront decomposition is particularly useful to offer a clear insight into the nature of the aberrations of an optical system.
Theoretical and experimental metalens aberration analysis and comparison with traditional lenses has been done in several works \cite{decker2019imaging,banerji2019imaging,khadir2021metasurface},
however, mainly for lenses with \emph{hyperbolic phase profiles}.

Here, we use the least squares method from Ref.~\cite{dai2008orthonormal} to decompose the wavefront of metalenses with \emph{quadratic phase profiles}, allowing for fast and boundary stable
calculations with orthonormal
basis even on noncircular apertures. The numerically simulated wavefront of the quadratic metalens was referenced to the perfect spherical wavefront originating from the point source,
numerically determined as the maximum intensity of the point spread function (PSF). 
The decomposition analysis was performed for two cases: metalens with a diameter $D=4f$ and $D=500\,\mu$m, both with a focal distance of $f=203\,\mu$m, which corresponds to our numerically simulated and experimentally fabricated metalenses.
As shown in the main text, the effective working area of the quadratic metalens is laterally shifted within the total metalens aperture as a function of the angle of incidence (AOI).
Therefore, Zernike decomposition was performed by setting the center as the center of the effective working area.

\begin{figure}[ht]
\centering\includegraphics[width=13cm]{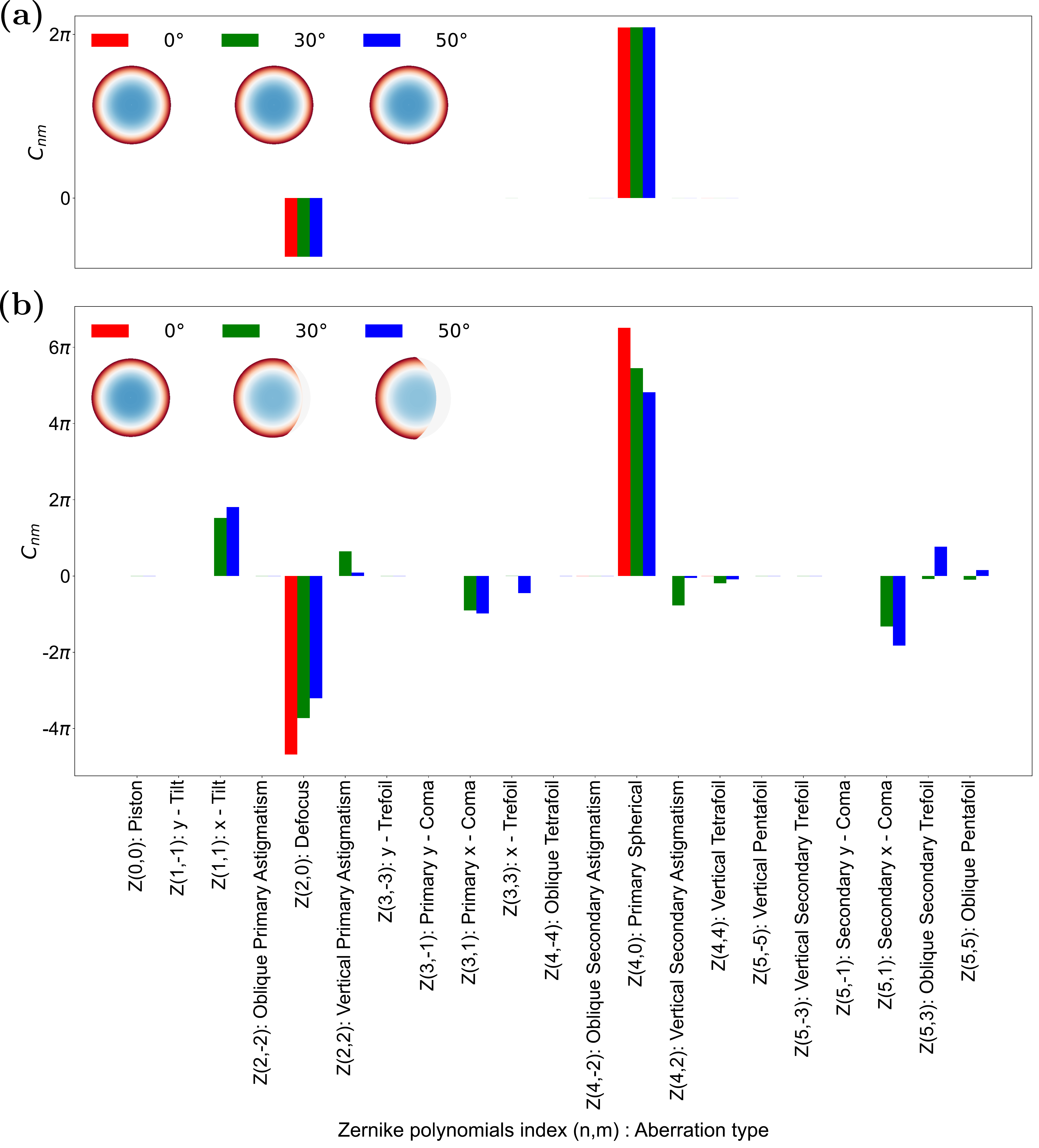}
\caption{Zernike decomposition of the wavefront of a quadratic metalens with the focal length $f=203\,\mu$m and diameter (a) $D=4f$ and (b) $D=500\,\mu$m for three different AOI: $\theta_\text{i}=0^\circ{}$, $30^\circ{}$ and $50^\circ{}$.
  The insets show the corresponding wavefronts referenced to perfect spherical profiles.}
\label{fig:Zernike}
\end{figure}

\begin{figure}[ht]
\centering\includegraphics[width=13cm]{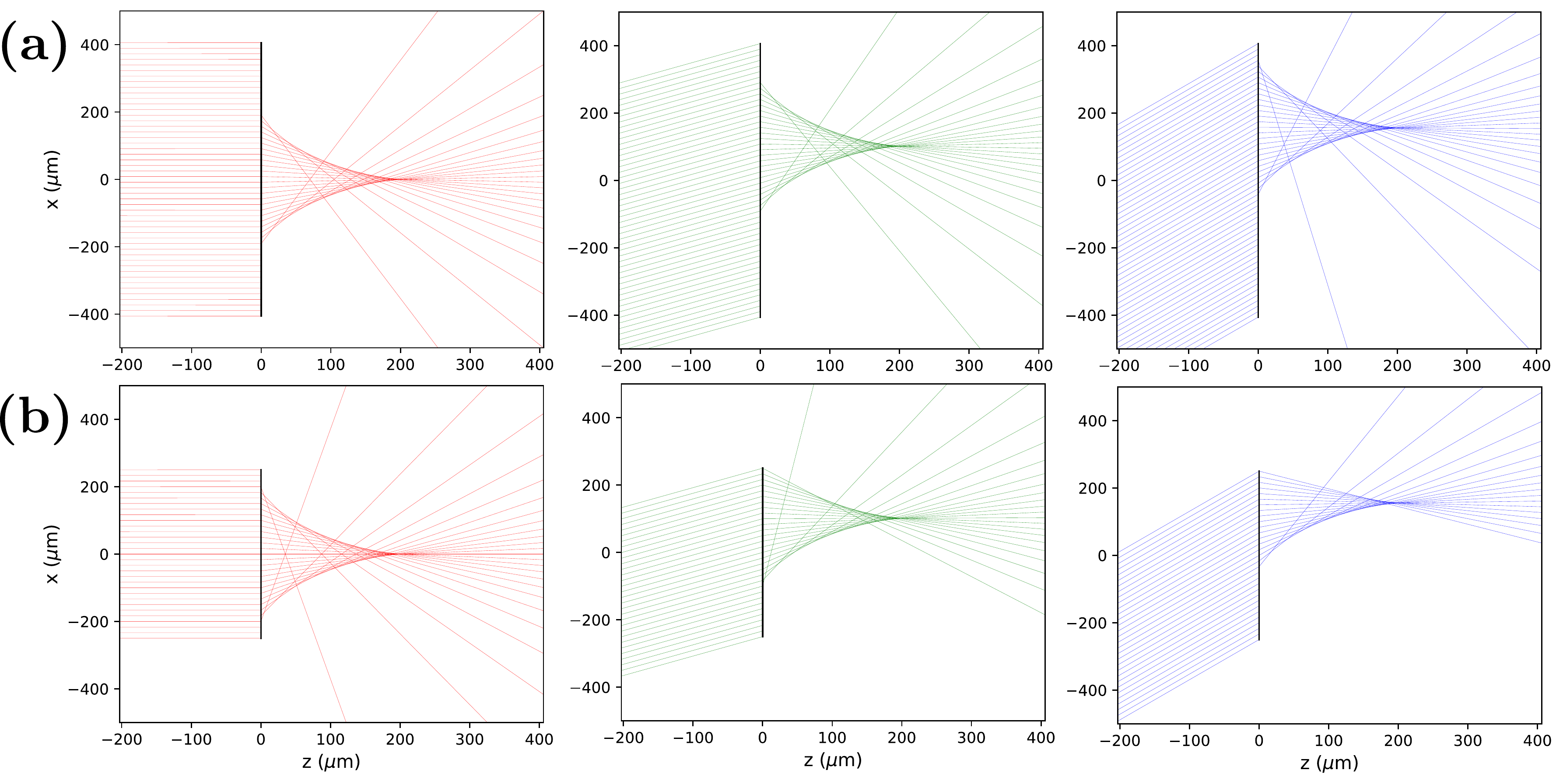}
\caption{Ray-tracing of a quadratic metalens of focal lens $f=203\,\mu$m with diameter (a) $D=4f$ and (b) $D=500\,\mu$m for three different AOI: $\theta_\text{i}=0^\circ{}$ (in red), $30^\circ{}$ (in green) and $50^\circ{}$ (in blue). The lens lies in
the plane $z=0$ (black solid line).}
\label{fig:ray-tracing_Ramon}
\end{figure}

For the first case ($D=4f$), the Zernike analysis for various AOI ($\theta_\text{i}=0^\circ{}$, $30^\circ{}$ and $50^\circ{}$) is shown in
Fig.~\ref{fig:Zernike}~(a).
In this case, the effective working area remains uncropped, fully inside the real aperture (\emph{i.e.} the physical size of the lens, see insets),
and the major aberrations are \emph{spherical} and \emph{defocus}.
Importantly, the value of low order decomposition coefficients is not affected by the AOI (at least up to an AOI of $50^\circ{}$ as investigated here). Hence, aberrations and focus quality remain unaffected, as observed in 
Fig.~1 of the main text. The spherical aberrations are clearly visible in the ray-tracing picture as shown in Fig.~\ref{fig:ray-tracing_Ramon}~(a), where the farther the rays are from the optical axis (OA),
the closer to the lens they intersect the OA (positive spherical aberration).
The results of the Zernike decomposition for the second case ($D=500\,\mu$m) are given in Fig.~\ref{fig:Zernike}~(b) as a function of the AOI.
One can see that the effective working area is cropped by the real aperture of the lens for large AOI (see insets). Hence, additional aberrations, \emph{e.g.} \emph{coma} and \emph{trefoil}, occur.
This is illustrated in Fig.~\ref{fig:ray-tracing_Ramon}~(b), where in addition to spherical aberrations, the coma aberrations become particularly visible at AOI $\theta_\text{i}=50^\circ{}$ (in blue).
Overall, in both cases, it is the spherical aberrations that dominate, followed by defocus aberrations, and by other ones like coma aberrations when the effective area starts to be cropped.

\newpage
\section{Focusing properties}

\begin{figure}[ht]
\centering\includegraphics[width=13cm]{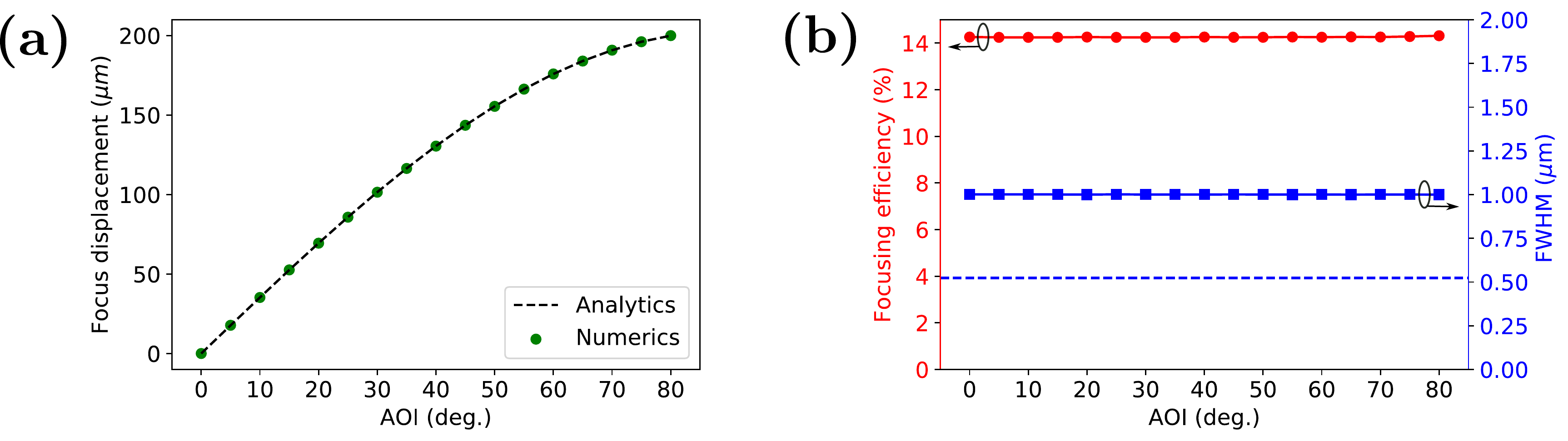}
\caption{Fourier optics simulations of the behaviour of a quadratic metalens. (a) Displacement of the focal spot in the lateral direction ($x$-direction) as a function of the AOI (green dots). The analytical formula $x(\theta_\text{i})=f\text{sin}(\theta_\text{i})$
  [see Eq.~(7) in the main text] is also shown (dashed black line). (b) Focusing efficiency (red dots) and FWHM of the focal spot (blue squares) as a function of the AOI.
  The FWHM corresponding to a diffraction-limited lens with NA$=\sqrt{2}/2$ is also shown (dashed blue line).}
\label{fig:fig1}
\end{figure}

\section{Imaging properties}
\label{sec_sup:imaging}

In this Section, we derive the ellipse equation for the image formation using ray-tracing methods.
We start by recalling the case of an \emph{ideal} thin lens, in the sense of a perfect stigmatic system that images a point source into a single image point.
For such a system, it is sufficient to consider two particular rays coming from an object point to obtain the corresponding image point as the intersection of these two rays in the image space.
This approach is also known as the \emph{first-order} ray-tracing method.

\begin{figure}[ht]
\centering\includegraphics[width=\textwidth]{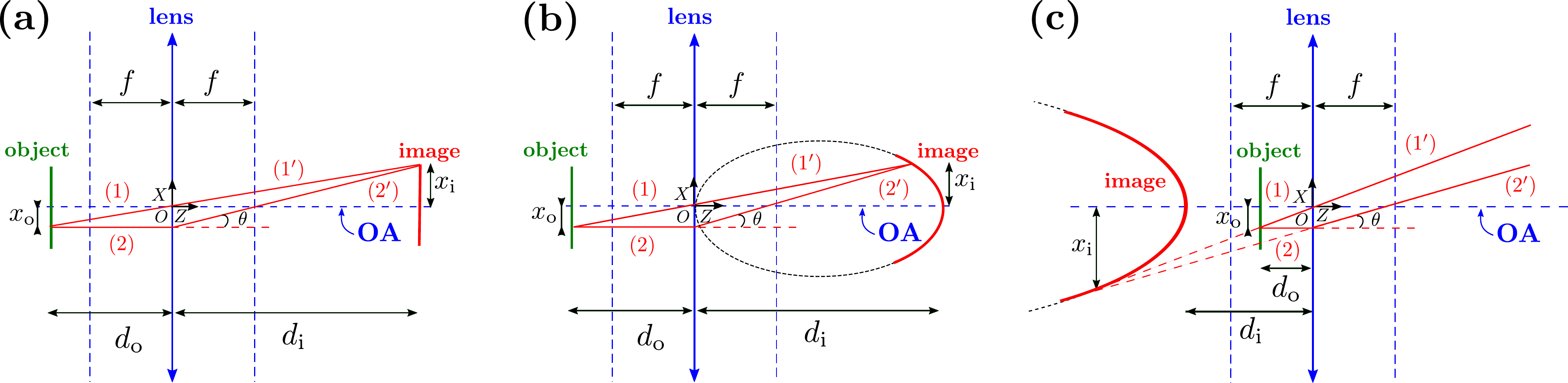}
\caption{First-order ray-tracing for (a) an ideal and (b), (c) a quadratic lens of focal length $f$ (represented by the blue, double-arrowed line).
  (a) In the ideal case, the construction of the image of a planar object (green line) located at a distance $d_\text{o}$
  from the lens is based on two special rays (red lines labeled by $(1)$, $(1')$, $(2)$ and $(2')$) starting from an object point $x_\text{o}$.
  Their intersection at the point $x_\text{i}$ gives
  the corresponding planar image (red line), at a distance $d_\text{i}$ from the lens given by Eq.~(\ref{eq:ideal_lenseq}). (b), (c) In the case of a quadratic lens, the intersection of the rays (red lines) coming
  from the planar object (green line)
  forms a curve image (red curve) which follows the equation of an ellipse given by Eq.~(\ref{eq:ellipse_image}) when $d_\text{o}>f$, or the equation of an hyperbola given by Eq.~(\ref{eq:ellipse_hyp}) when $d_\text{o}<f$ (dashed black lines). In configurations (a) and (b), $d_\text{o}=1.5f$ and $d_\text{i}=3f$, and in configuration (c), $d_\text{o}=0.6f$ and $d_\text{i}=1.5f$.}
\label{fig:ray_tracing_paraxial}
\end{figure}

For an ideal thin lens, one uses in general two particular rays, namely, the ray passing through the center of the lens [labeled $(1)$ in Figure~\ref{fig:ray_tracing_paraxial}~(a)],
and the ray parallel to the optical axis [labeled $(2)$ in Fig.~\ref{fig:ray_tracing_paraxial}~(a)]. The later will be deflected such that it passes through the focal point in the image space [labeled $(2')$ in Fig.~\ref{fig:ray_tracing_paraxial}~(a)], and the former will not be deflected [labeled $(1')$ in Fig.~\ref{fig:ray_tracing_paraxial}~(a)].
By considering a planar object located at a distance $d_\text{o}$ from the lens, and one point of this object at position $x_\text{o}$ (distance from optical axis),
the Rays (1) and (2) in the object space are defined by the following equations in the Cartesian coordinate system ($O$,$X$, $Z$) (all quantities being taken as algebraic quantities here):
\begin{align*}
  \text{Ray }(1):&\quad x=\frac{x_\text{o}}{d_\text{o}} z\\
  \text{Ray }(2):&\quad x=-x_\text{o}
\end{align*}
The rays in the image space are defined by the following equations:
\begin{align*}
  \text{Ray }(1'):&\quad x=\frac{x_\text{o}}{d_\text{o}} z\\
  \text{Ray }(2'):&\quad x=\text{tan}\left(\theta\right) z - x_\text{o}
%  \label{eq:ideal_x}
\end{align*}
where $\text{tan}\left(\theta\right)=x_\text{o}/f$.
The image point is found at the intersection of Rays $(1')$ and $(2')$ and reads:
\begin{equation}
  z=\frac{f d_\text{o}}{d_\text{o}-f}\; .
  \label{eq:ideal_lenseq}
\end{equation}
The last equation means that a planar object located at $d_\text{o}$ produces a planar image located at $d_\text{i}\equiv z=fd_\text{o}/(d_\text{o}-f)$, which is the thin lens equation usually given in the form: $1/d_\text{o}+1/d_\text{i}=1/f$ (see Fig.~\ref{fig:ray_tracing_paraxial}~(a)).
Moreover, by combining this thin lens equation with the equation of Ray $(1')$, one obtains the magnification: $x_\text{i}\equiv x=(d_\text{i}/d_\text{o})x_\text{o}$.
One can see that the size of an image linearly depends on the object size, which is a characteristic of an \emph{ideal} lens.
These conclusions are in general valid for real optical systems in the paraxial approximation, that is for object points close to the OA and small AOI ($x_\text{o}\ll d_\text{o}$).

\bigskip
Now, we apply this first-order ray-tracing method to the case a quadratic metalens.
For a quadratic lens, a ray parallel to the optical axis and reaching the lens at position $x$ is deflected by an angle $\theta$ which obeys: $\text{sin}\left(\theta\right)=x/f$.
This means that the equation of Ray $(2')$ must be amended by expressing $\text{tan}\left(\theta\right)$
as $\text{tan}\left(\theta\right)=\text{tan}\left[\text{arcsin}(x/f)\right]$.
Moreover, using the property that $\text{tan}\left[\text{arcsin}(x/f)\right]=x/f\times 1/\sqrt{1-(x/f)^2}$, the equation of Ray~$(2')$ becomes:
\begin{align*}
%  \label{eq:para_x}
  \text{Ray }(2'):\quad x=\frac{x_\text{o}/f}{\sqrt{1-\left(x_\text{o}/f\right)^2}} z - x_\text{o}
%  \label{eq:para_x_bis}
\end{align*}
The image point that is found at the intersection of Rays $(1')$ and $(2')$ now reads:
\begin{equation}
z+d_\text{o}=\frac{d_\text{o}}{f}\frac{1}{\sqrt{1-\left(\frac{d_\text{o}}{f}\frac{x}{z}\right)^2}} z   \; ,
\label{SIeq:add}
\end{equation}
where we made used of the equation of Ray $(1')$ to eliminate the $x_\text{o}$ variable and rearranged the terms.
In order to derive the ellipse equation, we make the paraxial approximation, that is $x_\text{o}\ll d_\text{o}$, which implies that $x\ll z$ according to the equation of Ray $(1')$. This allows us to make a Taylor expansion of the square-root term in Eq.~(\ref{SIeq:add}) as:
\begin{equation}
    \frac{1}{\sqrt{1-\left(\frac{d_\text{o}}{f}\frac{x}{z}\right)^2}}\simeq 1+\frac{1}{2}\left(\frac{d_\text{o}}{f}\frac{x}{z}\right)^2\; .
\end{equation}
Thus, after some simple algebraic manipulations, Eq.~(\ref{SIeq:add}) can be cast in the following forms, depending whether $d_\text{o}>f$ or $d_\text{o}<f$. In the case where $d_\text{o}>f$, the two Rays $(1')$ and $(2')$ intersect in the zone $z>0$ (real image) according to:
\begin{equation}
\frac{(z-z_0)^2}{a^2_\text{i}}+\frac{x^2}{b^2_\text{i}}=1\quad\text{with}\quad \left\{
    \begin{array}{ll}
       z_0=\frac{1}{2}\frac{d_\text{o}f}{d_\text{o}-f}  \\[2mm]
       a_\text{i}=\frac{1}{2}\frac{d_\text{o}f}{d_\text{o}-f} \\[2mm]
       b_\text{i}=\frac{f^2}{\sqrt{2d_\text{o}(d_\text{o}-f)}}
    \end{array}
    \right.\; ,
    \label{eq:ellipse_image}
\end{equation}
which is the ellipse equation given in the main text (see Fig.~\ref{fig:ray_tracing_paraxial}~(b)).
In the case where $d_\text{o}<f$, the two Rays $(1')$ and $(2')$ intersect in the zone $z<0$ (virtual image) according to:
\begin{equation}
\frac{(z+z_0')^2}{a_\text{i}'^2}-\frac{x^2}{b_\text{i}'^2}=1\quad\text{with}\quad \left\{
    \begin{array}{ll}
       z_0'=\frac{1}{2}\frac{d_\text{o}f}{f-d_\text{o}}  \\[2mm]
       a_\text{i}'=\frac{1}{2}\frac{d_\text{o}f}{f-d_\text{o}} \\[2mm]
       b_\text{i}'=\frac{f^2}{\sqrt{2d_\text{o}(f-d_\text{o})}}
    \end{array}
    \right.\; ,
    \label{eq:ellipse_hyp}
\end{equation}
which is the equation of a hyperbola centered on $(0,-z_0')$ with transverse axis length $2a_\text{i}'$ and conjugate axis length $2b_\text{i}'$ (see Fig.~\ref{fig:ray_tracing_paraxial}~(c)).

\section{Criterion for field-of-view} 

\begin{figure}[ht]
	\centering\includegraphics[width=8cm]{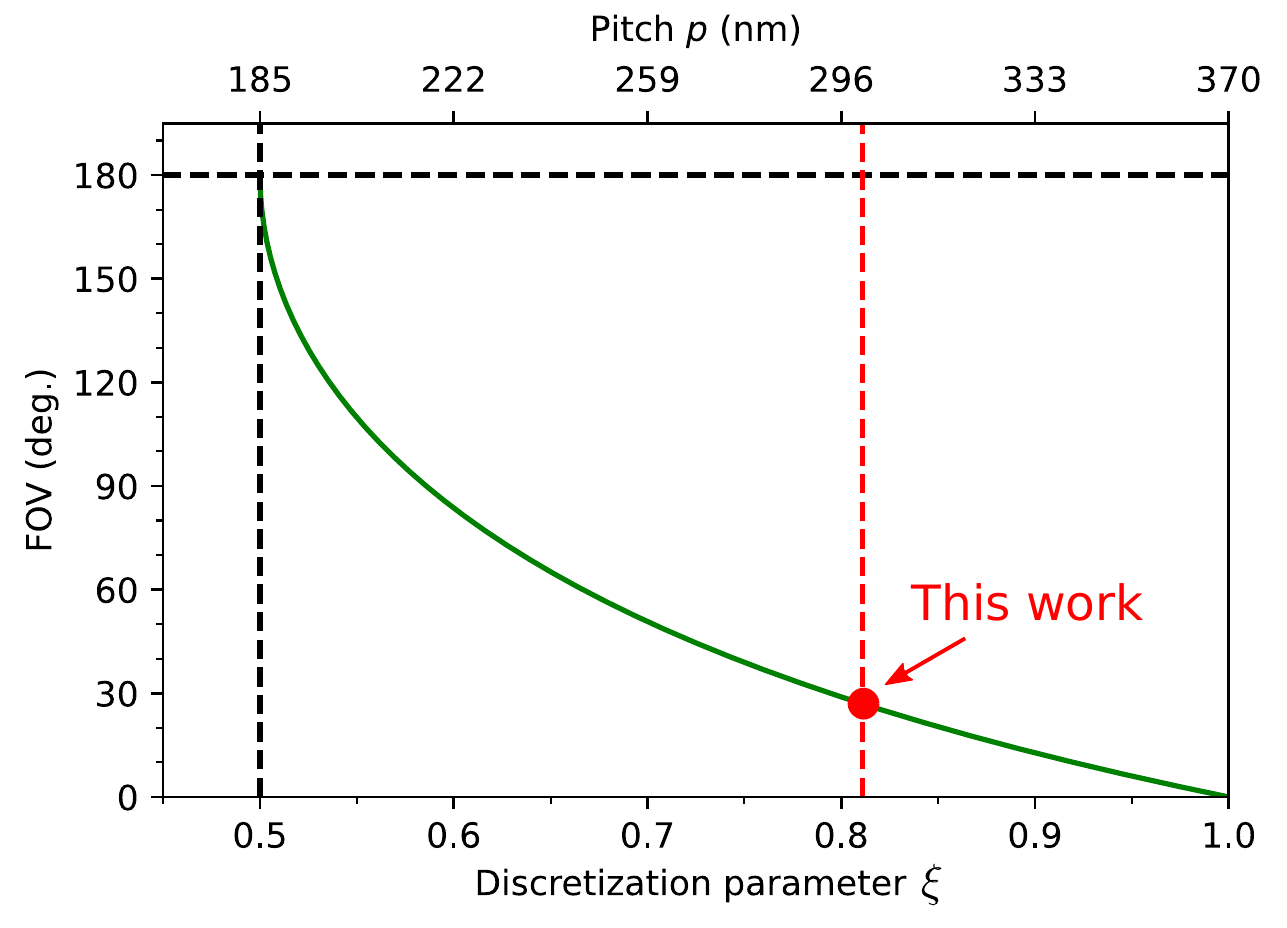}
	\caption{Minimum FOV of a quadratic metalens as given by Eq.~(18) in the main text as a function of the discretization parameter $\xi$ (lower axis) and the pitch $p$ (upper axis).
    The later is plotted for $\lambda_0=740\,$nm and for the minimum unit-cell number $N=2$, which corresponds to our fabricated metalens.}
\label{fig:figS4bis}
\end{figure}

\section{Angle of incidence study}
\label{sec_sup:design}

\begin{figure}[ht]
\centering\includegraphics[width=13cm]{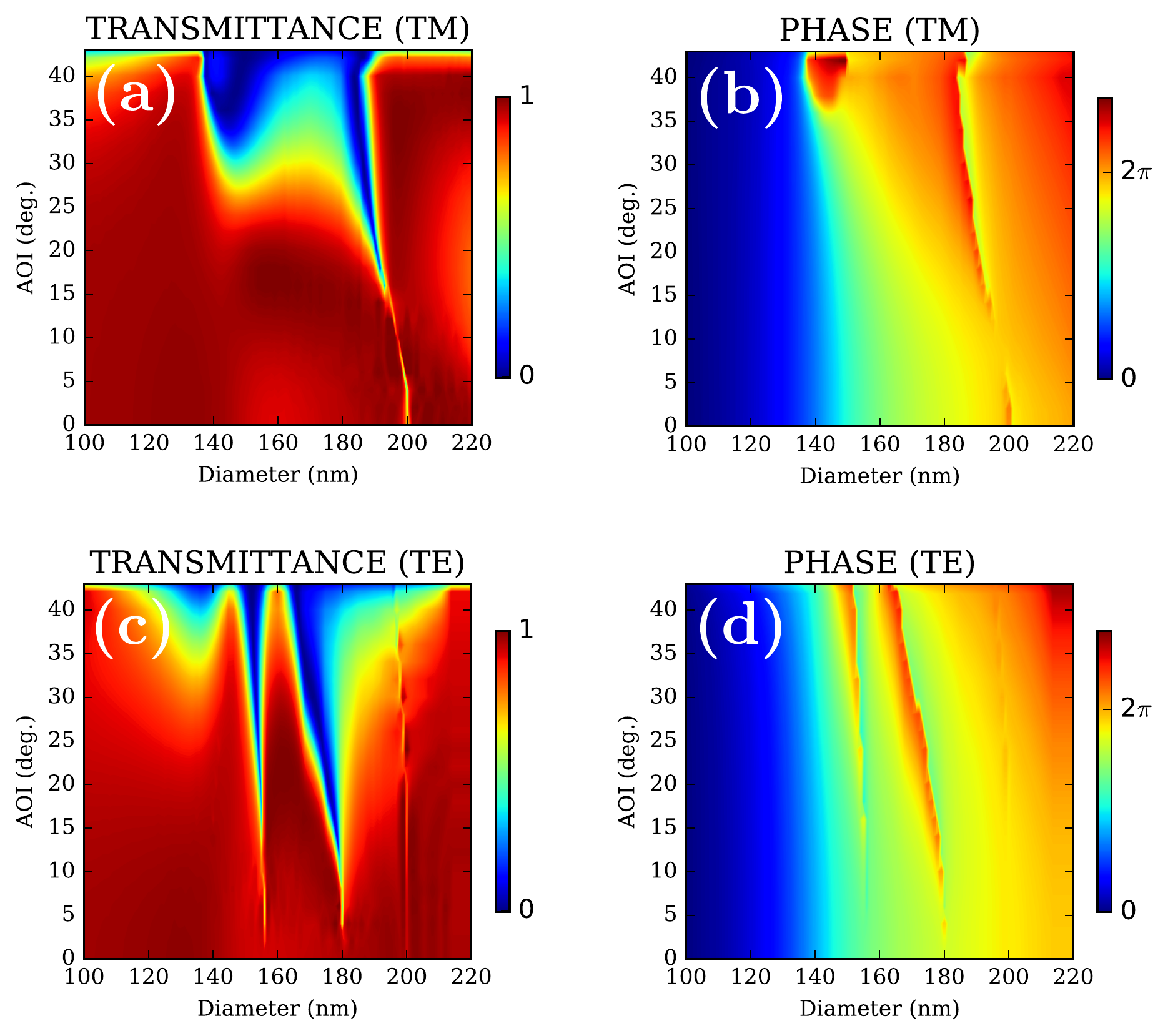}
\caption{Transmission efficiencies (transmittances) and phases of a periodic arrangement of identical nanopillars as a function of their diameters and AOI.
  (a), (b) Case of a TM incident wave. (c), (d) Case of a TE incident wave.}
\label{fig:trans_and_phase_maps}
\end{figure}

In addition to the simulations of the transmittance and phase as a function of the nanopillars diameters (shown in Fig.~3~(a) of the main text),
we also simulated the influence of the AOI on these quantities.
The results are shown in Fig.~\ref{fig:trans_and_phase_maps} as a function of the diameters and AOI, for a transverse magnetic (TM) incident wave in Fig.~\ref{fig:trans_and_phase_maps}~(a) and (b), respectively,
and for a transverse electric (TE) incident wave in Fig.~\ref{fig:trans_and_phase_maps}~(c) and (d), respectively.
One can see that the phase vary only weakly with the AOI, while the transmittance remain very high for most of the AOI and diameter parameters, except in some parameter regions
where Fabry-Perot resonances occur and are responsible for strong reflections (for example, for diameters around $153\,$nm for the TE wave).

Note that we only show AOI up to $\theta_\text{i}=43^\circ{}$, which corresponds to the critical angle for an interface glass ($n=1.46$) - air ($n=1$);
beyond this critical angle, the transmission efficiency is zero, due to the total internal reflection. However, this effect is a peculiarity of simulations that consider periodic arrangements of \emph{identical} nanopillars, and does not occur in the metalens since the \emph{gradient} of phase
imparted through \emph{different} nanopillars gives the required extra momentum for light to be transmitted.
To show that, we also simulated the transmission efficiencies of four different phase-gradient metasurfaces made of supercells with respectively $N=9$, $6$, $3$ and $2$ nanopillars,
each nanopillars being encompassed in a ``unit-cell'' of cross section $p\times p$ with $p=300\,$nm.
Each supercell samples linearly the $2\pi$ phase by its number of element $N$ by carefully choosing the correct nanopillars (see first column in Table~\ref{tab_sup:grating}).
These four supercells are representative of the ones used in the quadratic metalens.
In Table~\ref{tab_sup:grating}, we show the computed transmission efficiencies (in red) for TM and TE incident waves and for three different AOI $\theta_\text{i}=0^\circ{}$, $30^\circ{}$ and $50^\circ{}$,
together with the relative efficiencies into the blazed diffraction order (in blue).
One can see that energy is still transmitted at $\theta_\text{i}=50^\circ{}$ (which is above the critical angle), for the reasons given above.
The fact that there is no energy into the blazed order at $\theta_\text{i}=0^\circ{}$ in the case of supercells with $N=2$ elements (second column / last row in Table~\ref{tab_sup:grating}) is because such a surpercell is sub-diffractive at normal incidence
(its length $N p<\lambda_0$ with $\lambda_0=740\,$nm).
Note, however, that for the AOI $\theta_\text{i}=30^\circ{}$ and $50^\circ{}$ the first-order (blazed) diffraction channel becomes propagating and the supercell becomes ``active''. In particular, for certain configurations of polarizations and AOI, only the channel corresponding to the blazed order is available, and therefore this channel has 100\% of relative efficiency (see Table~\ref{tab_sup:grating}).

\newpage

\begin{table}
\centering
\begin{tabular}{|c|c|c|c|}
\hline
$N$ & $0^\circ{}$ (TM / TE) & $30^\circ{}$ (TM / TE) & $50^\circ{}$ (TM / TE) \\
\hline
\multirow{2}{*}{\begin{minipage}{.3\textwidth}
      \centering \includegraphics[width=\linewidth, height=10mm]{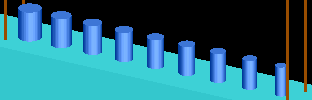}
    \end{minipage}} & \textcolor{red}{$97\%$ / $94\%$} & \textcolor{red}{$59\%$ / $82\%$} & \textcolor{red}{$28\%$ / $40\%$} \\
& \textcolor{blue}{$95\%$ / $98\%$} & \textcolor{blue}{$37\%$ / $55\%$} & \textcolor{blue}{$86\%$ / $28\%$} \\
\hline
\multirow{2}{*}{\begin{minipage}{.25\textwidth}
      \centering \includegraphics[width=\linewidth, height=10mm]{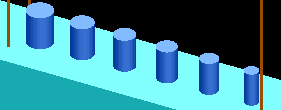}
    \end{minipage}} & \textcolor{red}{$92\%$ / $94\%$} & \textcolor{red}{$56\%$ / $80\%$} & \textcolor{red}{$35\%$ / $44\%$} \\
& \textcolor{blue}{$93\%$ / $96\%$} & \textcolor{blue}{$44\%$ / $90\%$} & \textcolor{blue}{$77\%$ / $31\%$} \\
  \hline
  \multirow{2}{*}{\begin{minipage}{.18\textwidth}
      \includegraphics[width=\linewidth, height=8mm]{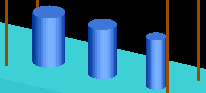}
    \end{minipage}} & \textcolor{red}{$45\%$ / $63\%$} & \textcolor{red}{$73\%$ / $77\%$} & \textcolor{red}{$33\%$ / $26\%$} \\
  & \textcolor{blue}{$61\%$ / $74\%$} & \textcolor{blue}{$67\%$ / $76\%$} & \textcolor{blue}{$100\%$ / $71\%$} \\
    \hline
      \multirow{2}{*}{\begin{minipage}{.12\textwidth}
      \includegraphics[width=\linewidth, height=8mm]{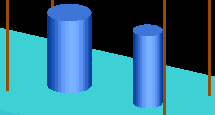}
    \end{minipage}} & \textcolor{red}{$14\%$ / $62\%$} & \textcolor{red}{$67\%$ / $97\%$} & \textcolor{red}{$38\%$ / $59\%$} \\
  & \textcolor{blue}{$0\%$ / $0\%$} & \textcolor{blue}{$70\%$ / $99\%$} & \textcolor{blue}{$100\%$ / $100\%$} \\
\hline
\end{tabular}
\caption{Transmission efficiencies (in red) for different AOI $\theta_\text{i}=0^\circ{}$, $30^\circ{}$ and $50^\circ{}$ of different phase-gradient metasurfaces containing supercells with $N=9$, $6$, $3$ and $2$ elements, respectively, computed for TM and TE incident waves. The relative efficiencies into the blazed diffractive order are also shown (in blue).}
  \label{tab_sup:grating}
\end{table}

\newpage
\section{Barrel distortion}

In this Section, we derive analytical formulas to quantify the barrel distortion effect, in order to unveil the existing trade-off between the maximum FOV of an object that can be imaged and the optical resolution of the metalens and/or the detector resolution.
The barrel distortion, or fish-eye effect, is the fact that equally spaced object points produce image points that get more compressed in the edges of the image. More specifically,
one can show that, for an object point located at a distance $x_\text{o}$ from the OA (see Fig.~4~(a) in the main text), a quadratic metalens will produce an image point located at $x_\text{i}$:
\begin{equation}
x_\text{i}=f\,\frac{x_\text{o}/d}{\sqrt{1+\left(x_\text{o}/d\right)^2}}\; .
  \label{eq:Xi}
\end{equation}

\begin{figure}[ht]
\centering\includegraphics[width=13cm]{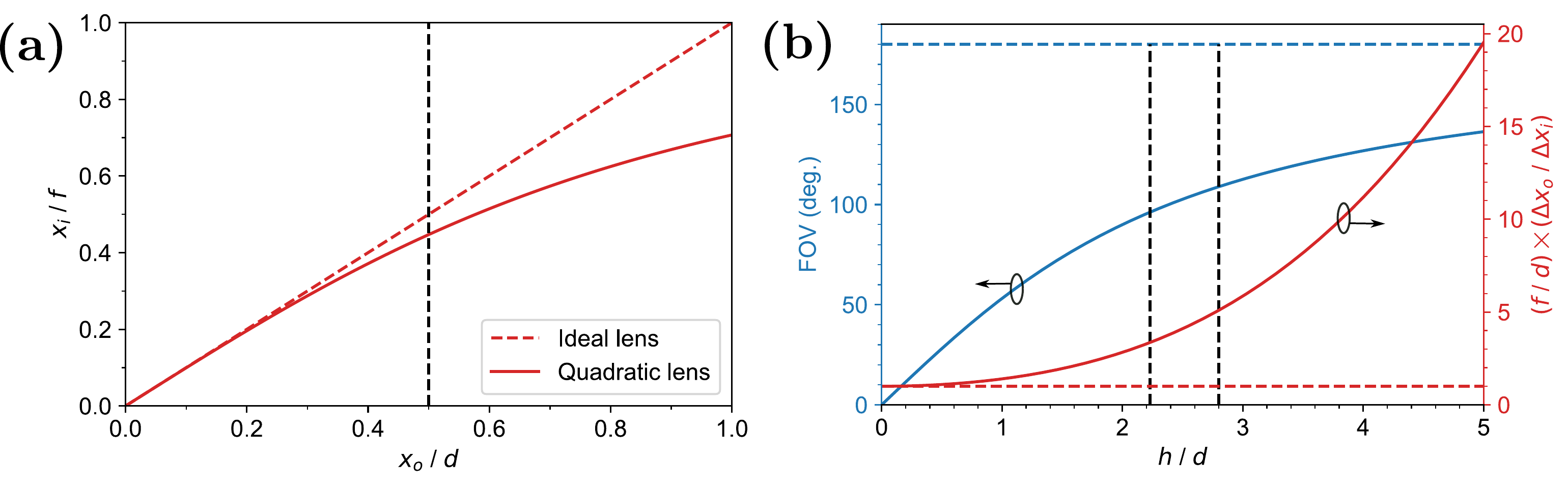}
\caption{Barrel distortion effect. (a) Point-to-point imaging: image point $x_\text{i}$ normalized by $f$ as a function of the corresponding object point $x_\text{o}$ normalized by $d$ for a quadratic lens
  (solid red line). The case of an ideal lens is also shown (dashed red line).
  (b) Relative image resolution $\Delta x_\text{o}/\Delta x_\text{i}$ normalized by $d/f$ (solid red line) and object FOV (solid blue line) as a function of the object size $h$ normalized by $d$.
  The limit FOV $=180^\circ{}$ is also shown (dashed blue line). The relative resolution in the case of an ideal lens $\Delta x_\text{o}/\Delta x_\text{i}= d/f$ is also shown (dashed red line). The two ruler imaging configurations \{$d=3\,$cm, $h=6.7\,$cm\} and \{$d=2\,$cm, $h=5.6\,$cm\} are indicated by the two vertical dashed lines (left and right, respectively).}
\label{fig:barrel_distortion}
\end{figure}

This behavior is shown in Fig.~\ref{fig:barrel_distortion}~(d), where we plot the image position given by Eq.~(\ref{eq:Xi}) as a function of the object position (solid red line),
together with the behavior of an \emph{ideal} lens with no distortion (dashed red line).
As seen, there is less than 10\% of distortion for object points located at a distance $x_\text{o}<0.5 d$ (this limit is shown by the vertical dashed line), which is barely visible to the naked eye.
Indeed, one can check by looking at Figs.~5~(b) and (c) in the main text, that those parts of the ruler up to $h=d$, corresponding to $h=3\,$cm and $h=2\,$cm, respectively, are almost not distorted.

One can see from Eq.~(\ref{eq:Xi}) that, as the distance of the object points $x_\text{o}$ increases, the distance $x_\text{i}$ increases at an increasingly slower pace (due to the denominator).
This in turn will increase the image spatial resolution required to resolve these points.
To quantify this, we start by differentiating Eq.~(\ref{eq:Xi}) with respect to the distance $x_\text{o}$ to obtain:
\begin{equation}
\frac{\Delta x_\text{o}}{\Delta x_\text{i}}=\frac{d}{f}\left[1+\left(\frac{x_\text{o}}{d}\right)^2\right]^{3/2}\; ,
 \label{eq:criteriabis}
\end{equation}
where $\Delta x_\text{o}$ is the distance between two adjacent object points to be resolved, and $\Delta x_\text{i}$ is the distance between the corresponding image points.
In practice, $\Delta x_\text{o}$ is the minimum object feature size that one wants to image and,
depending on which one is the limiting factor, $\Delta x_\text{i}$ corresponds to either
the detector spatial resolution, or to the optical spatial resolution of the lens.
In the following, we call the ratio $\Delta x_\text{o}/\Delta x_\text{i}$ the \emph{relative image resolution}.
In Fig.~\ref{fig:barrel_distortion}~(e), we plot (solid red line) this relative image resolution $\Delta x_\text{o}/\Delta x_\text{i}$ given by Eq.~(\ref{eq:criteriabis})
as a function of the object size $h$ ($x_\text{o}=h/2$) normalized by the distance $d$.
We also show the FOV of the object given by Eq.~(17) from the main text (solid blue line). One can see that, as the ratio $h/d$ increases, both the relative image resolution and FOV increase.

\begin{figure}[ht]
	\centering\includegraphics[width=8cm]{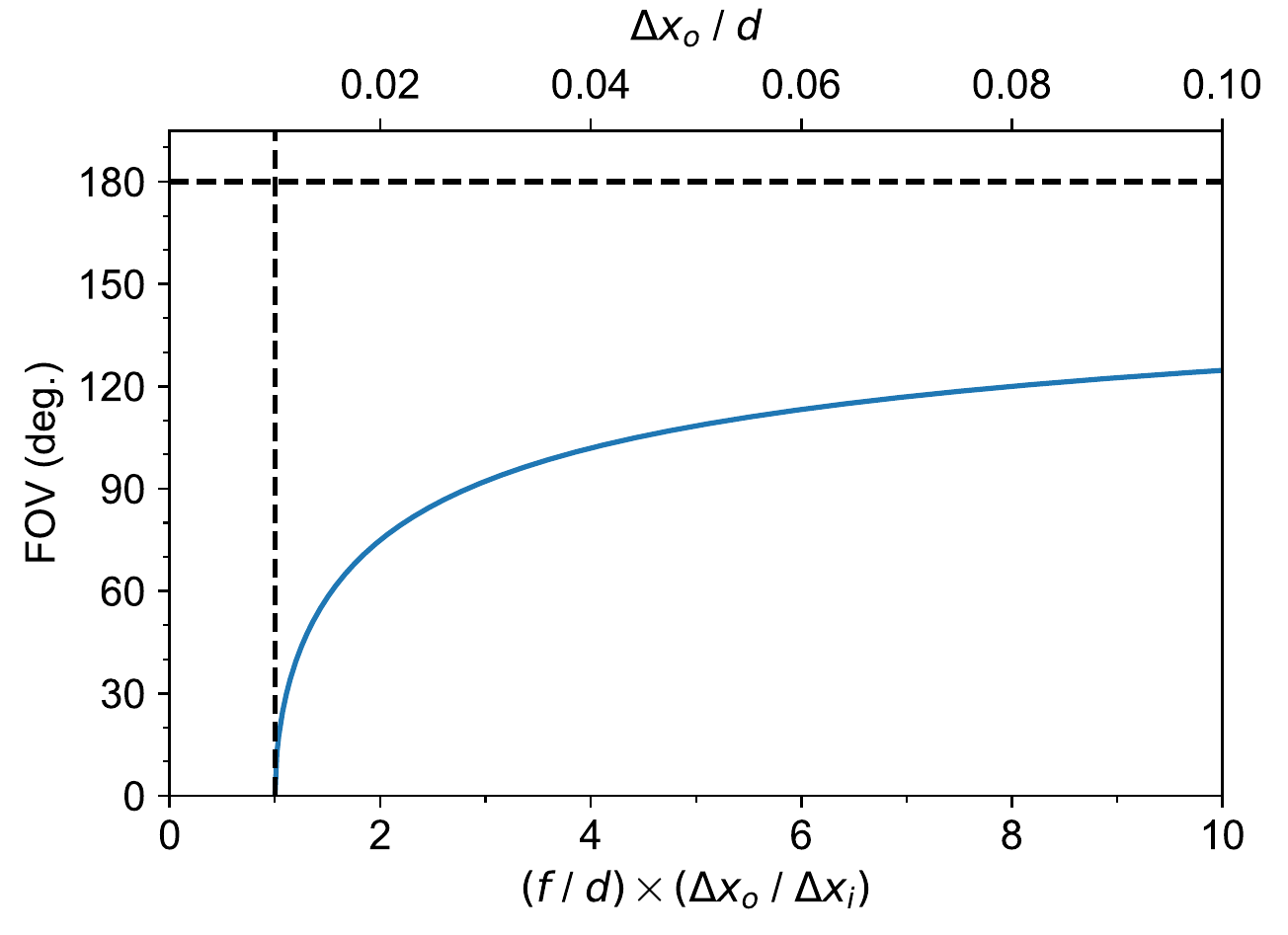}
	\caption{Maximum FOV of an object as given by Eq.~(\ref{eq:FOV_vs_res}) as a function of the relative spatial resolution normalized by $d/f$ (lower axis) and the ratio $\Delta x_\text{o}/d$ (upper axis). The later is plotted for $\Delta x_\text{i}=2\,\mu$m and $f=203\,\mu$m, which corresponds to our fabricated metalens.}
\label{fig:figS6}
\end{figure}

Finally, by using Eq.~(17) from the main text into Eq.~(\ref{eq:criteriabis}), we find the relation between the FOV of an object that can be resolved and the relative spatial resolution $\Delta x_\text{o}/\Delta x_\text{i}$:
\begin{equation}
  \text{FOV}=2\,\text{arctan}\left(\sqrt{\left(\frac{f}{d}\frac{\Delta x_\text{o}}{\Delta x_\text{i}}\right)^{\frac{2}{3}}-1}\right)\; ,
  \label{eq:FOV_vs_res}
\end{equation}  
which corresponds to Eq.~(18) of the main text.
In other words, this equation predicts, for a given image spatial resolution $\Delta x_\text{i}$, what is the maximum FOV of an object with features down to $\Delta x_\text{o}$ in size that can be resolved
(see Fig.~\ref{fig:figS6} for a plot of this maximum FOV). The application of this formula to the ruler case
in the configurations $d=3\,$cm and $d=2\,$cm gives, with $\Delta x_\text{o}\sim 1\,$mm and $\Delta x_\text{i}\sim 2\,\mu$m, maximum FOV of $96^\circ{}$ and $109^\circ{}$, respectively.

\section{Imaging experiment}

\begin{figure}[ht]
	\centering\includegraphics[width=13cm]{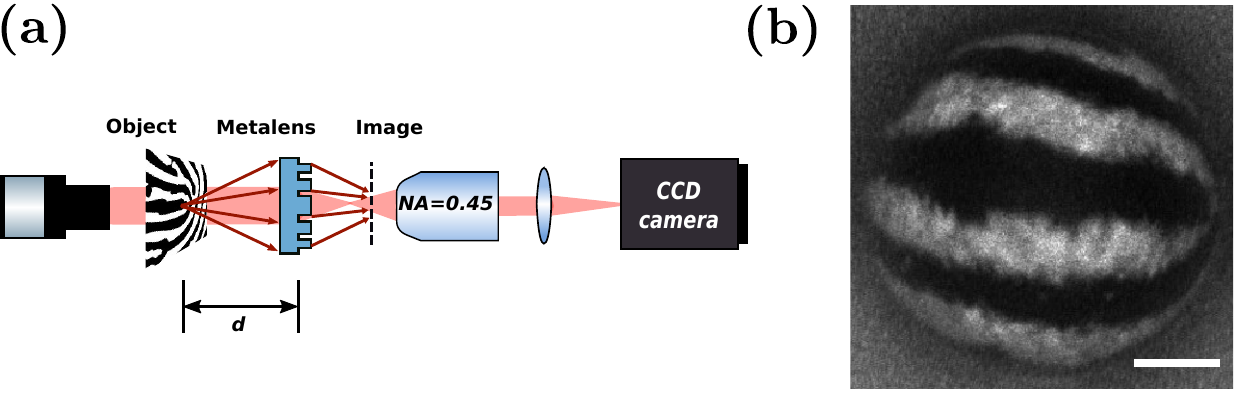}
	\caption{Imaging experiment. (a) Schematic of the optical setup. The light red color denotes the laser beam path in the absence of the object.
          (b) The result of the experimental imaging of the fingerprint at $d=800\,\mu$m. The scale bar corresponds to $50\,\mu$m.
	}
\label{fig:figS4}
\end{figure}

The main text presents the imaging of the ruler and the fingerprint. Both were printed on a white paper and illuminated by the laser in a transmission configuration.
Since the ruler is an elongated object, a Thorlabs $0.4^{\circ}\times 100^{\circ}$ line engineered diffuser was utilized to produce a line pattern.
The image produced by the metalens was transferred to a CCD camera by a free-space microscope setup with 5X magnification (20X Nikon infinity-corrected objective, NA=0.45,
a tube lens with 50 mm focal length, see Fig.~\ref{fig:figS4}~(a)). 

As far as the fingerprint imaging is concerned, as discussed in the main text, the smallest possible object-lens distance $d=800\,\mu$m corresponds to the thickness of the metalens substrate.
To demonstrate this most extreme case, we fabricated a small focal length metalens ($f=100\,\mu$m, $D=246\,\mu$m) and glued the fingerprint image on the opposite side of the substrate.
The imaging result is presented in Fig.~\ref{fig:figS4}~(b). As one can expect, the object size one can image is quite limited due to the barrel distortions, resulting in
$\sim 1.5\,\text{mm}\times 1.5\,$mm visible fingerprint area.  

\section{MTF analysis}

To give more insight on the quadratic lens imaging performance, we provide here a monochromatic modulation transfer function (MTF) calculation. The MTF of an optical system describes the image contrast attenuation arising from the finite resolution of the optics. Mathematically, the MTF is calculated as the normalized Fourier transform of the PSF, showing the contrast drop for each spatial frequency.

\begin{figure}[ht]
\centering\includegraphics[width=13cm]{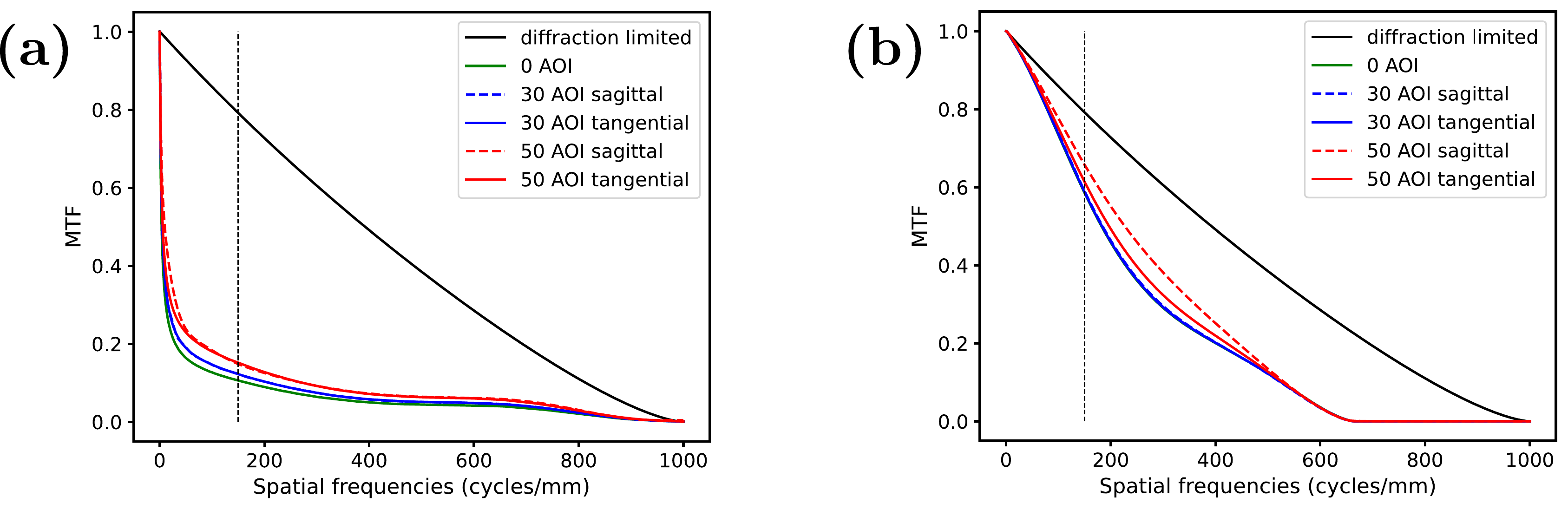}
\caption{MTF calculations for monochromatic illumination at $\lambda_0=740\,$nm for two different configurations: quadratic metalens (a) without and (b) with a pinhole of diameter $100\,\mu$ located at the front focal plane of the lens. The calculations are performed for three different AOI: $0^\circ{}$ (green line), $30^\circ{}$ (blue lines) and $50^\circ{}$ (red lines), and for both tangential and sagittal directions. The MTF of a diffraction-limited system with an $\text{NA}= 0.37$ is also shown (solid black line). The vertical black dashed line is a guide-to-the eye at $150$ cycles/mm.}
\label{fig:mtf}
\end{figure}

In Fig.~\ref{fig:mtf}~(a), we calculated the MTF at the working wavelength $\lambda_0=740\,$nm for on-axis ($0^\circ{}$ AOI, solid green line) and off-axis points ($30^\circ{}$ and $50^\circ{}$ AOI, blue and red lines, respectively) for both horizontal and vertical directions (called tangential and sagittal directions in the legend of the figure, respectively). The quadratic lens MTF is compared to a diffraction-limited system with an $\text{NA}= 0.37$ (solid black line; see also main text, Section 3.2.3). One can see that the quadratic lens exhibits a rapid contrast drop down to about $10\%$ at $150$ cycles/mm (vertical dashed black line), which is mainly associated with inherent spherical aberrations discussed in the main text.
One can also see that the cutoff spatial frequency (\emph{i.e.}, the spatial frequency above which the MTF is equal to zero), defined as $1000\times 2\text{NA}/\lambda_0$ (to have it in cycles/mm), is about the same for both and equal to about $1000$ cycles/mm.

To improve the contrast, an aperture stop (pinhole) located at the front focal plane of the lens can be utilized \cite{engelberg2020good}. In Fig.~\ref{fig:mtf}~(b), we present the MTF calculation for a pinhole-metalens system (the green and blue lines can barely be distinguished because they coincide almost perfectly). The pinhole has a diameter of $100\,\mu$m (\emph{i.e.}, about $f/2$). A glass substrate is taken as the medium located between the pinhole and the metalens. One can see that, while the cutoff spatial frequency in the pinhole-metalens system case is about 650 cycles/mm, due to the overall NA reduction, such an arrangement can provide a substantial improvement of the MTF, that may be beneficial for the realization of better devices.

%%%%%%%%%%%%%%%%%%%%%%% References %%%%%%%%%%%%%%%%%%%%%%%%%
%\pagebreak
%%%%%%%%%% If using BibTeX:
\bibliography{biblio_supp_mat}

%%%%%%%%%% If preparing manually:
% \newcommand{\enquote}[1]{``#1''}

% \bibitem{Zhang:14}
% Y.~Zhang, S.~Qiao, L.~Sun, Q.~W. Shi, W.~Huang, L.~Li, and Z.~Yang,
%   \enquote{Photoinduced active terahertz metamaterials with nanostructured
%   vanadium dioxide film deposited by sol-gel method,}
%   {\protect\JournalTitle{Optics Express}} \textbf{22}, 11070--11078 (2014).

% {Optical Society}, \enquote{{OSA Publishing},}
%   \url{http://www.osapublishing.org}.

% \bibitem{FORSTER2007}
% P.~Forster, V.~Ramaswamy, P.~Artaxo, T.~Bernsten, R.~Betts, D.~Fahey,
%   J.~Haywood, J.~Lean, D.~Lowe, G.~Myhre, J.~Nganga, R.~Prinn, G.~Raga,
%   M.~Schulz, and R.~V. Dorland, \enquote{Changes in atmospheric consituents and
%   in radiative forcing,} in \enquote{Climate Change 2007: The Physical Science
%   Basis. Contribution of Working Group 1 to the Fourth assesment report of
%   Intergovernmental Panel on Climate Change,}  S.~Solomon, D.~Qin, M.~Manning,
%   Z.~Chen, M.~Marquis, K.~B. Averyt, M.~Tignor, and H.~L. Miler, eds.
%   (Cambridge University Press, 2007).

% \end{thebibliography}